\documentclass[twocolumn,showpacs,preprintnumbers,amsmath,amssymb]{revtex4-1}
\usepackage{amsmath,amssymb,color,epsfig,latexsym,natbib,graphicx,dcolumn}
\usepackage{bm}
\bibpunct{[}{]}{,}{n}{,}{,}		
  
\def\DEL#1{{\textcolor{green}{}}}         

\def\ie{{\it i.e. }}

\def\rms{{\it r.m.s. }}

\newcommand{\bB}{\bf{B}}

\newcommand{\rem}[1]{}

\newcommand\vecp[1]{\vec{#1}}			

\newcommand{\be}{\begin{equation}}
\newcommand{\ee}{\end{equation}}

\def\bB0{\vecp{B}_0} 
\def\kpe{k_{\perp}} 
\def\pa{\parallel} 
\def\kpa{k_{\parallel}}

\begin{document}

\title{\bf On the two-dimensional state in driven magnetohydrodynamic turbulence} 

\author{Barbara Bigot$^{1,2}$ and S\'ebastien Galtier$^{3,4}$}
\affiliation{1. Space Science Center, University of New Hampshire, Durham, NH $03824$, USA}
\affiliation{2. Center for Integrated Computation and Analysis of Reconnection And Turbulence}
\affiliation{3. Univ Paris-Sud, Institut d'Astrophysique Spatiale, b\^at. 121, F-91405 Orsay, France}
\affiliation{4. Institut universitaire de France}

\date{\today}

\begin{abstract}
The dynamics of the two-dimensional (2D) state in driven tridimensional (3D) incompressible 
magnetohydrodynamic turbulence is investigated through high-resolution direct numerical simulations 
and in the presence of an external magnetic field at various intensities. For such a flow the 2D state (or 
slow mode) 
and the 3D modes correspond respectively to spectral fluctuations in the plan $k_\parallel=0$ and in the 
area $k_\parallel>0$. It is shown that if initially the 2D state is set to zero it becomes non negligible in few 
turnover times particularly when the external magnetic field is strong. The maintenance of a large scale 
driving leads to a break for the energy spectra of 3D modes; when the driving is stopped 
the previous break is removed and a decay phase emerges with alfv\'enic fluctuations. For a strong 
external magnetic field the energy at large perpendicular scales lies mainly in the 2D state and in 
all situations a pinning effect is observed at small scales. 
\end{abstract}

\maketitle

\section{Introduction}

A variety of astrophysical plasmas is well described by magnetohydrodynamic (MHD) turbulence in 
the compressible or even simply in the incompressible case \cite{elmegreen,Govoni,galtier06}. 
The solar wind is often cited as an example since many in situ data are available which have shown 
a medium characterized by turbulent fluctuations over a large band of frequency $f$ from a fraction of 
mHz to kHz. Whereas the turbulence regime at $f>1$Hz requires the introduction of non MHD processes 
like dispersion \cite{Galtier06aa,Galtier08a} the low frequency part at $f<1$Hz is the source of a lot of 
activities on MHD \cite{MontgoTurner,Grappin83,GS95,ng96,Amitava,politano98,Galtier99,Cho00,
Galtier2000,Galtier2002,Galtier2005,Mininni05,Muller,Boldyrev06,Galtier06,Alexakis07a,Alexakis07b,Bigot07a,
Bigot08b,Boldyrev08,Chandran08,Boldyrev09,Galtier09,Matt09,Mininni09,Muller10,Podesta10}. 

The first description proposed by Iroshnikov and Kraichnan (IK) \cite{iro,Kraichnan65} 
is a heuristic model for incompressible MHD turbulence \`a la Kolmogorov where the 
large-scale magnetic field is supposed to act on small-scales as a uniform magnetic field, leading 
to counterpropagating Alfv\'en wave packets whose interactions with turbulent motions produce a 
slowdown of the nonlinear energy cascade. The typical transfer time through the scales is then 
estimated as $\tau_{nl}^2/\tau_A$ (instead of $\tau_{nl}$ for Navier-Stokes turbulence), where 
$\tau_{nl} \sim \ell /u_{\ell}$ is the eddy turnover time at characteristic length scale $\ell$ and 
$u_{\ell}$ is the associated velocity. The Alfv\'en time is the time of collision between counterpropagating 
wave packets and is estimated as $\tau_A \sim \ell/B_0$ where $B_0$ represents the large-scale magnetic
field normalized to a velocity (${\bf B_0} \to {\bf B_0} \sqrt{\mu_0\rho_0}$). (This renormalization will be
used in the rest of the paper.) Hence, the energy spectrum in $k^{-3/2}$ unlike the $k^{-5/3}$ Kolmogorov 
one for neutral flows. The weakness of the IKs phenomenology is the apparent contradiction between the 
presence of Alfv\'en waves and the absence of an external uniform magnetic field: the external field is 
supposed to be played by the large-scale magnetic field but its main effect -- \ie anisotropy -- is not included 
in the description. 

Two fundamental evolutions in MHD turbulence have been made during the last two decades. There are 
both concerned with anisotropy. The first one is the conjecture that the refined times 
$\tau_{nl} \sim \ell_\perp / u_{\ell_\perp}$ and $\tau_{A} \sim \ell_\parallel / B_0$ (where $\perp$ and 
$\parallel$ are respectively the perpendicular and parallel directions to the mean magnetic field ${\bf B_0}$) 
are balanced at all scales \cite{GS95}. It leads to the heuristic $k_\perp^{-5/3}$ energy spectrum as well as 
the relationships $\kpa \sim \kpe^{2/3}$. Whereas the first relation is a trivial consequence of the conjecture 
(and often wrongly interpreted as the main result of the conjecture) the second prediction reveals a non trivial 
character of MHD turbulence. The second fundamental evolution is the possibility to handle the effects of a 
strong $B_0$ on the MHD dynamics through a rigorous mathematical treatment of weak turbulence which 
leads asymptotically to a set of integro-differential equations. The exact solution for weak turbulence at zero 
cross-helicity is a $k_\perp^{-2}$ energy spectrum \cite{Galtier2000}. Note that the form of the 
energy spectrum in the regime of strong turbulence is still the subject of discussions 
\cite{Galtier2005,Boldyrev06} although the relationships $\kpa \sim \kpe^{2/3}$ seems to be often verified, 
whereas the weak turbulence prediction has been obtained recently by two independent set of direct 
numerical simulations \cite{Bigot08b,Boldyrev08}. 

The transition from strong to weak turbulence has been the subject of a previous paper where the regime 
of decaying turbulence has been investigated \cite{Bigot08b}. In the present paper the focus is made on 
the regime of driven incompressible MHD turbulence under the influence of a uniform magnetic field at 
various intensity and in the balance case. A set of high-resolution direct numerical simulations in the 
tridimensional case are reported. Particular attention is turned to the dynamics of the 2D state (also called 
slow mode) which 
corresponds by definition to the fluctuations in the plan $k_\parallel=0$. In particular, we demonstrate the 
fundamental role of the 2D state at large $B_0$. 

The organization of the paper is as follows. In Section \ref{sec2} the numerical setup is given as well as 
the definition of the 3D modes {\it versus} the 2D state. In Section \ref{sec4}, the temporal characteristics 
of the different flows are investigated whereas the study of the energy spectra is exposed in Section 
\ref{sec5}. Finally, a summary and a conclusion are given in the last Section.

\section{Numerical setup}
\label{sec2}
\subsection{MHD equations}

The incompressible MHD equations in the presence of a uniform magnetic field ${\bf B_0}$ read 
\begin{eqnarray}
\partial_t {\bf v} - B_0 \partial_{\parallel} {\bf b} + {\bf v} \cdot \nabla \, {\bf v} &=& 
- {\bf \nabla} P_* + {\bf b} \cdot \nabla \, {\bf b} + \nu \Delta {\bf v} \, , 
\label{mhd1} \\
\partial_t {\bf b} - B_0 \partial_{\parallel} {\bf v} + {\bf v} \cdot \nabla \, {\bf b} &=& 
{\bf b} \cdot \nabla \, {\bf v} + \eta \Delta {\bf b} \, , 
\label{mhd2} \\
\nabla \cdot {\bf v} &=& 0 \, , 
\label{mhd1b} \\
\nabla \cdot {\bf b} &=& 0 \, ,
\label{mhd2b}
\end{eqnarray}
where ${\bf v}$ is the velocity, ${\bf b}$ the magnetic field (in velocity unit), $P_*$ the total 
(magnetic plus kinetic) pressure, $\nu$ the viscosity and $\eta$ the magnetic diffusivity. 
The introduction of the Els\"asser fields ${\bf z}^\pm ={\bf u} \pm {\bf b}$ for the fluctuations is also 
useful; it gives (assuming a unit magnetic Prandtl number, \ie $\nu=\eta$)
\begin{eqnarray}
\partial_t {\bf z}^\pm \mp B_0 \partial_{\parallel} {\bf z}^\pm + {\bf z}^\mp  \cdot \nabla {\bf z}^\pm &=& 
-\nabla P_* + \nu \nabla^2 {\bf z}^\pm  \, , 
\label{mhd3} \\
\nabla \cdot {\bf z}^\pm &=& 0 \, . 
\label{mhd4}
\end{eqnarray}
The third term in the left hand side of Eq.~(\ref{mhd3}) represents the nonlinear interactions between the 
${\bf z}^\pm$ fields, while the second term represents the linear Alfv\'enic wave propagation along the 
${\bf B_0}$ field which defines the z-direction. Note that both descriptions (\ref{mhd1})--(\ref{mhd2b}) and 
(\ref{mhd3})--(\ref{mhd4}) will be used through the paper.

\begin{table*}
\caption{\label{table1}
Computational parameters are given for runs {\bf Ia} to {\bf IIIa} for which initially $E^v=E^b$ and for run 
{\bf IIIb} for which initially $E^v>E^b$ (see text). Spatial resolution, viscosity $\nu$ ($\nu=\eta$) and magnetic 
field intensity $B_0$ are given, followed by the initial values of the integral length scales (perpendicular 
$L_\perp= 2\pi\int{(E^v(\kpe) /  \kpe) d\kpe}/\int{E^v(\kpe) d\kpe}$ and parallel 
$L_\parallel= 2\pi\int{(E^v(\kpa)/\kpa)d\kpa}/\int{E^v(\kpa)d\kpa}$), the \rms velocity 
$u_{rms}=<{\bf v}^2>^{1/2}$, the \rms magnetic field $b_{rms}=<{\bf b}^2>^{1/2}$, the reduced cross-helicity $\rho$, 
the kinetic Reynolds number $\mathcal{R}_v=\ u_{rms}L_\perp/\nu$ (which is equal to the magnetic Reynolds 
number), the eddy turnover time $\tau_{nl}=L_\perp/u_{rms}$ and the Alfv\'en time $\tau_{A}=L_\parallel/B_0$.  
All these quantities are also computed in the stationary phase for which turbulence is fully developed. 
Note that the value of $\tau_A$ for run {\bf Ia} in the stationary phase is overestimated since $b_{rms} > B_0$.) 
We finally give the time $t^*$ from which the driving is stopped and $t_{M}$ the final time of the simulation.}
\begin{ruledtabular}
\begin{tabular}{cc|c|c|c}
&  $\hspace*{20mm} $  & $\rm{Initial \hspace*{2mm} conditions \hspace*{25mm}}$ & $\rm{Stationary\hspace*{2mm}phase\hspace*{32mm}}$ & $ $ \\ 
\end{tabular}
\begin{tabular}{cccc|cccccccc|cccccccc|cc}
&  $  $ & $\nu$ & $B_0$ & $L_\perp$ & $L_\parallel$ & $u_{rms}$& $b_{rms}$ & $\rho$ & $\mathcal{R}_v$ & $\tau_{nl}$ & $\tau_A$ & 
                          $L_\perp$ & $L_\parallel$ & $u_{rms}$& $b_{rms}$ & $\rho$ & $\mathcal{R}_v$ & $\tau_{nl}$ & $\tau_A$ & 
                          $t^{*}$ & $t_{M}$ \\
{\bf Ia} & $512^3$ & $2.10^{-3}$ & $1$ &$3.45$ & $4.95$ & $1$ & $1$ & $0$ & $1775$ & $3.45$& $4.96$ 
                                       &$2.1$ & $3.8$ & $1.8$ & $2.3$ & $0.05$--$0.12$ &$1890$ & $1.17$& ${\it 3.8}$
                                       &$7$ & $11$ \\
{\bf IIa} & $512^3$ & $2.10^{-3}$ & $5$ &$3.45$ &$4.95$ & $1$ &$1$ & $0$ & $1775$ & $3.45$ & $0.99$ 
                                       &$2.4$ & $5$ & $1.9$ & $2$ & $0.2$--$0.3$ &$2280$ & $1,26$& $1$
                                       &$9$ & $12$ \\
{\bf IIIa} & $512^3$ & $2.10^{-3}$ & $15$ &$3.45$ &$4.95$ & $1$& $1$ & $0$ & $1775$ & $3.45$ & $0.33$ 
                                       &$2.4$ & $5.7$ & $1,7$ & $2.1$ & $0.3$ & $2040$ & $1,41$& $0,38$
                                       &$6$ & $9$ \\
{\bf IIIb}  & $512^3$ & $2.10^{-3}$ & $15$ & $3.49$ & $4.92$ & $1.16$ & $0.80$& $0$ &$2024$ & $3$ & $0.33$ 
                                       &$2.6$ & $5.6$ & $1.8$ & $2.1$& $0.1$ &$2340$ & $1.44$& $0.37$
                                       &$5.1$ & $6$ \\
\end{tabular}
\end{ruledtabular}
\end{table*}

\subsection{Initial conditions}
\label{IC}

We numerically integrate the 3D incompressible MHD equations (\ref{mhd1})--(\ref{mhd2b}) in a 
$2\pi$-periodic box, using a massively parallel pseudo-spectral code including de-aliasing and with 
a spatial resolution of $512 \times 512 \times 512$ grid-points. The time marching uses an 
Adams-Bashforth / Cranck-Nicholson scheme which is a second-order finite-difference scheme 
in time \citep{ponty}. A unit magnetic Prandlt number is chosen \ie $\nu=\eta$ for all runs. 
We present below the different set of simulations.

\subsubsection{Runs Ia to IIIa}

The first set of numerical simulations is characterized by an external force which fixes the bidimensional 
Els\"asser spectra as 
\be
E^\pm(\kpe,\kpa) = F(\kpa) \kpe^3 \, , 
\label{f1}
\ee
where only the two first perpendicular and parallel wavenumbers are energized with $F(1) = F(2) = 1/2$. 
The 2D state, $E^\pm(\kpe,0)$, is never forced which means that initially it has no energy and may evolve 
freely at time $t>0$. 

The kinetic
\be
E^v = {1 \over 2} <{\bf u}^2({\bf x})> \, ,
\ee
and magnetic 
\be
E^b = {1 \over 2} <{\bf b}^2({\bf x})> \, ,
\ee 
energies are initially equal such that $E^v=E^b=1/2$. Note that $< \cdot >$ means a space averaging.

The reduced cross-helicity between the velocity and magnetic field fluctuations which is measured by 
\be
\rho \equiv {2<{\bf u}({\bf x}) \cdot {\bf b}({\bf x)}> \over <{\bf u}^2({\bf x}) + {\bf b}^2({\bf x})>} \, ,
\label{rho}
\ee
is initially set to zero. 
The initial (large-scale) kinetic and magnetic Reynolds numbers are about $1775$ for the flows with 
$\nu= 2 \times 10^{-3}$ (see Table~\ref{table1}), with $u_{rms}=b_{rms}=1$. 

The perpendicular integral length is  
\be
L_\perp = 2\pi {\int{ (E^v(\kpe)/\kpe) d\kpe} \over \int{E^v(\kpe) d\kpe}} \sim 3.5  \, , 
\ee
whereas the parallel integral length is
\be
L_\pa = 2\pi {\int{ (E^v(\kpa)/\kpa) d\kpa} \over \int{E^v(\kpa) d\kpa}} \sim 5 \, .
\ee

A parametric study is performed according to the intensity of $B_0$. Three different values are used, 
namely $B_0 = 1, 5$ and $15$. All these simulations are driven up to time $t^*$ for which a state 
of fully developed turbulence is reached. Then, the driving is stopped and the flows evolve freely. 
These simulations correspond respectively to runs ${\bf Ia}$ to ${\bf IIIa}$ (see Table~\ref{table1}).

\subsubsection{Run IIIb}

The second type of simulation is characterized by an external force on the bidimensional Els\"asser 
spectra such that 
\be
E^\pm(\kpe,\kpa) = F(\kpa) \kpe^2 \, , 
\ee
where $F$ is the same as in (\ref{f1}). The kinetic energy is initially larger than the magnetic energy with 
$E^v \simeq 2 \times E^b=0.67$ and only the case $B_0=15$ is considered. 

All information presented in this Section is summarized in Table~\ref{table1}. Additionally, we give the 
computational parameters when the stationary phase is reached for which a balance is obtained between 
forcing and dissipation.

\subsection{3D modes and $2$D state}
\label{sec3}

In the presence of an external magnetic field ${\bf B}_0$, it is convenient to describe the flow dynamics 
in terms of Alfv\'en waves which propagate along ${\bf B}_0$ at frequency 
$\omega({\bf k})={\bf k} \cdot {\bf B}_0=k_\parallel B_0$. In such a description, we may define the 3D modes, 
namely the spectral fluctuations at $k_\pa \neq 0$, for which the Alfv\'en frequency is non-zero. The 
complementary part namely the spectral fluctuations at $k_\pa=0$ defines the 2D state (or slow mode). 
Note that the 2D state can still be associated to waves since a local mean magnetic field ${\bf b_0}$ may be 
defined in the $k_\pa=0$ plan along which local Afv\'en waves propagate with frequency 
$\omega_{2D} ({\bf k})= {\bf k}_\perp\cdot {\bf b}_0$. 

To obtain the fields associated to the 3D modes and the 2D state a simple decomposition is performed 
from the Fourier space. For the 3D modes, the Els\"asser, velocity and magnetic fields are defined as 
${\bf \hat{z}}^{\pm}(k_\perp,k_\pa>0)$, ${\bf \hat{u}}(k_\perp,k_\pa>0)$ and ${\bf \hat{b}}(k_\perp,k_\pa>0)$ 
respectively. For the $2$D state, the velocity and magnetic fields are defined as ${\bf \hat{u}}(k_\perp,k_\pa=0)$ 
and ${\bf \hat{b}}(k_\perp,k_\pa=0)$. From these quantities, we define the Els\"asser fields of the 2D state as  
\be 
{\bf \hat{z}}^\pm(k_\perp,k_\pa=0) = {\bf \hat{u}}(k_\perp,k_\pa=0) \pm  {\bf \hat{b}}(k_\perp,k_\pa=0) \, .  
\ee
In the rest of the paper, the quantities associated to the 3D modes and $2$D state will be noted with ``$w$'' 
and ``$2D$'' indices respectively. Note that a different notation was used in \cite{Bigot08b} where shear- 
and pseudo-Alfv\'en waves were defined from a toroidal/poloidal decomposition.

\section{Temporal analysis}
\label{sec4}

In this Section, we study the temporal behavior of several global quantities to characterize the MHD 
flow dynamics and the influence of the $B_0$ intensity. In all the following Figures, time evolutions 
are shown for simulations {\bf Ia} to {\bf IIIa} and {\bf IIIb} (see Table~\ref{table1}). 

\begin{figure}[ht]
\resizebox{86mm}{!}{\includegraphics{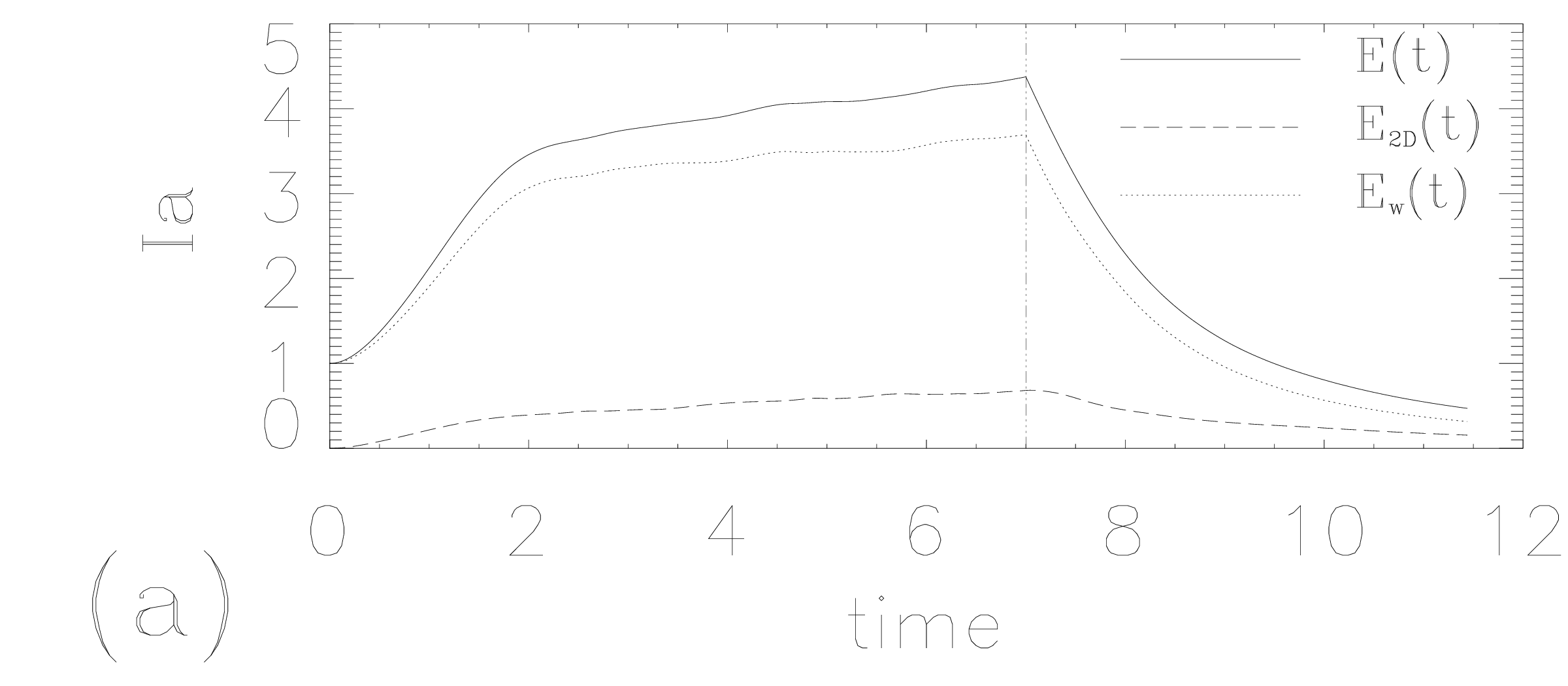}}\\
\resizebox{86mm}{!}{\includegraphics{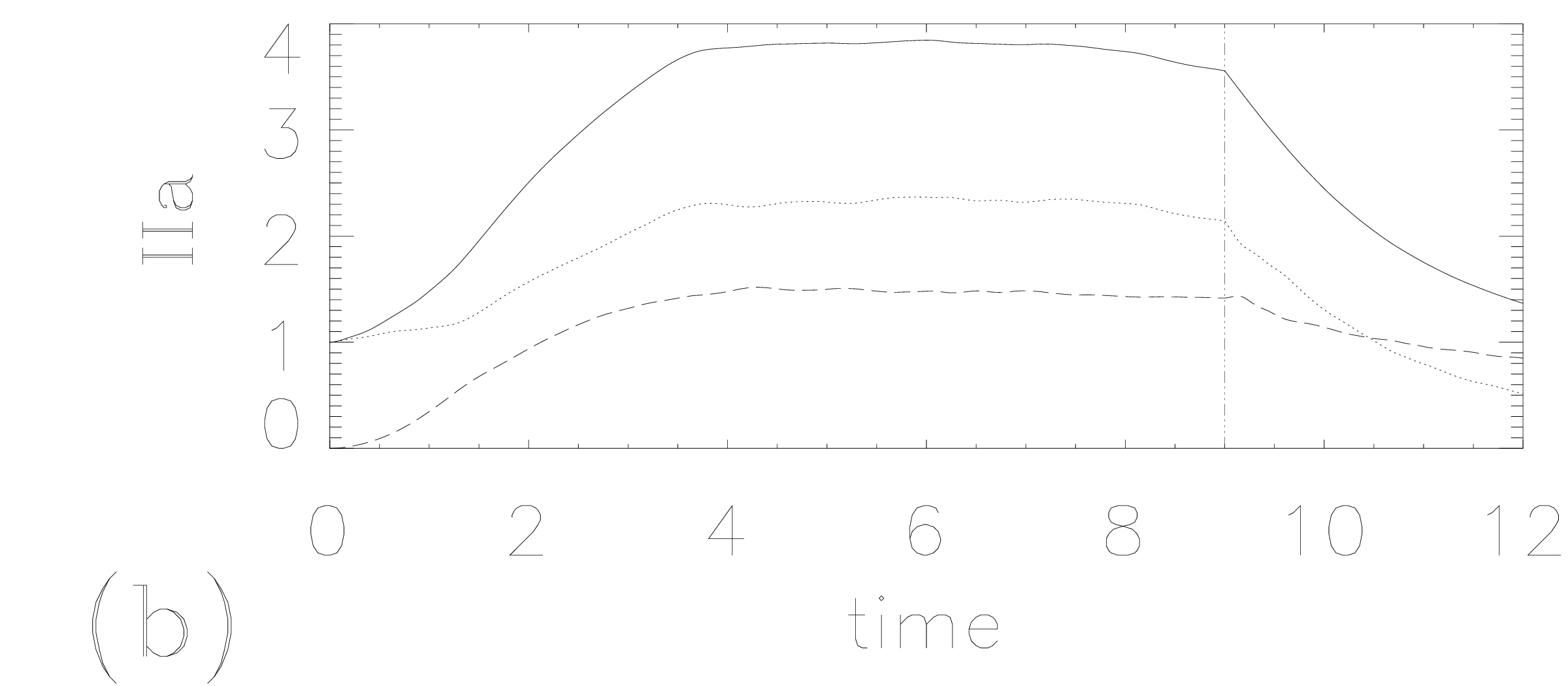}}\\
\resizebox{86mm}{!}{\includegraphics{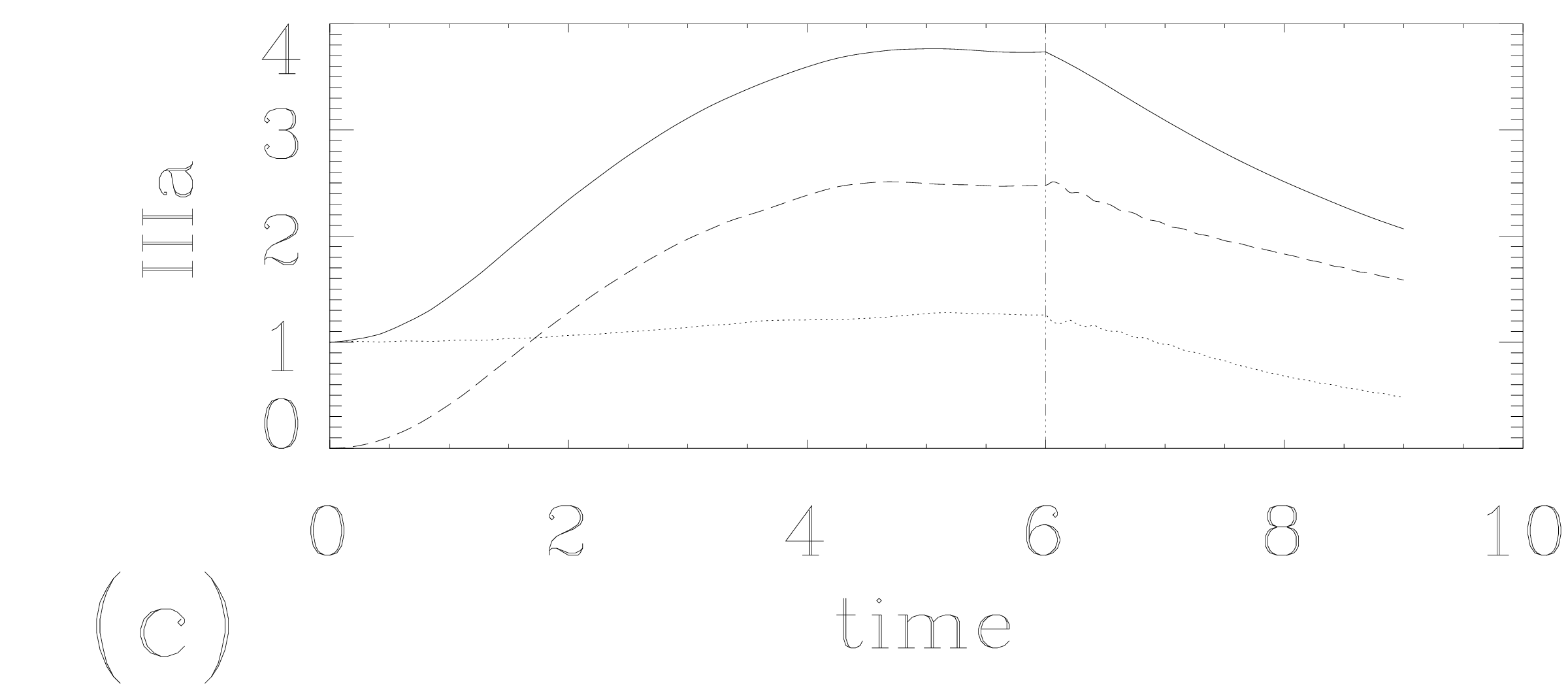}}
\caption{Temporal evolution of $E(t)$ (solid), $E_{2D}($t$)$ (dash) and $E_w($t$)$ (dot) for $B_0=1$ (a), 
$5$ (b) and $15$ (c) (runs ${\bf Ia}$ to ${\bf IIIa}$). Note that the vertical lines correspond to $t^*$. 
\label{FigtmpEb0}}
\end{figure}

We first consider the evolutions of the total energy 
\be
E(t) = E^+(t)+E^-(t) \, , 
\ee 
where 
\be 
E^\pm(t) = \frac{1}{2} \int_{\kpa} \int_{k_\perp}{\bf \hat{z}}^{\pm^2}(k_\perp,k_\pa) dk_\perp d\kpa \, ,
\label{deft}
\ee
the energy of the 3D modes 
\be
E_w(t) = E_w^+(t) + E_w^-(t) \, ,  
\ee
where
\be 
E_w^\pm(t) = \frac{1}{2} \int_{\kpa>0} \int_{k_\perp}{\bf \hat{z}}^{\pm^2}(k_\perp,k_\pa) dk_\perp d\kpa \, ,
\label{defw}
\ee
and the energy of the $2$D state
\be
E_{2D}(t) = E(t) - E_w(t) \, .
\label{def2d}
\ee
The temporal evolution of these energies are displayed in Fig.~\ref{FigtmpEb0}. For all simulations, 
we distinguish three parts in the evolution of the total energy $E(t)$. The first part is characterized by an 
increase of the total energy which corresponds to a non stationary phase where the balance between the 
external force and the dissipation is not reached. When it is reached a plateau is observed: it is the second 
phase. Finally, a third phase appears when the external force is suppressed at time $t^*$ (see 
Table~\ref{table1}). Then, a decrease of the total energy is found during this decay phase. 
Note that the second phase is reached later when the mean magnetic field is stronger; it is compatible with 
a simple heuristic analysis in terms of time scales which shows that a transfer time in $\tau_{nl}^2 / \tau_A$ is 
indeed larger at larger $B_0$. 

The main difference observed between these three simulations comes from the evolution of the energies 
$E_w(t)$ and $E_{2D}(t)$. We remind that initially there is no energy in the 2D state, but as we can see, 
shortly after the beginning of the simulations the 2D state is excited. For the case $B_0=1$, the energy 
increases mainly in the 3D modes whereas the energy of the 2D state reaches in the stationary phase 
approximatively $16\%$ of the total energy. A different behavior is found for the case $B_0=15$ where a 
strong increase of the energy of the 2D state is measured which reaches in the stationary phase around 
$2/3$ of the total energy. In this case the energy of the 2D state is significantly larger than the one of the 
3D modes which displays only a weak variation during the driving phase. 
The case $B_0=5$ is an intermediate case where the energy of the 2D state stabilizes approximatively 
at $1/3$ of the total energy.

It is important to note that during the first phase the increase of $b_{rms}$ leads to a decrease 
of the ratio $B_0/b_{rms}$ to around $0.4$, $2.5$ and $7.1$ for runs ${\bf Ia}$, ${\bf IIa}$ and ${\bf IIIa}$ 
respectively, with the characteristic timescale relations $\tau_A \sim \tau_{nl}$ for runs ${\bf I-IIa}$ and 
$\tau_A \ll \tau_{nl}$ for runs ${\bf IIIa}$ (see Table~\ref{table1}). This evolution has an impact on the 
nonlinear dynamics as we can see in run ${\bf IIa}$: the energy of the 3D modes has a weak variation until 
time $t \sim1.5$ for which we have $B_0/b_{rms} \sim 3.3$. After this period of time a stronger increase of the 
energy of the 3D modes is noted until a stationary phase is reached. The value $B_0/b_{rms} \sim 3.3$ appears 
to be a threshold beyond which the energy in the 3D modes is roughly conserved. This analysis is confirmed by 
run ${\bf IIIa}$ where the energy of the 3D modes does not change significantly during the period of driving 
for which we always have $B_0/b_{rms} > 3.3$. 

\begin{figure*}[ht]
\resizebox{88mm}{!}{\includegraphics{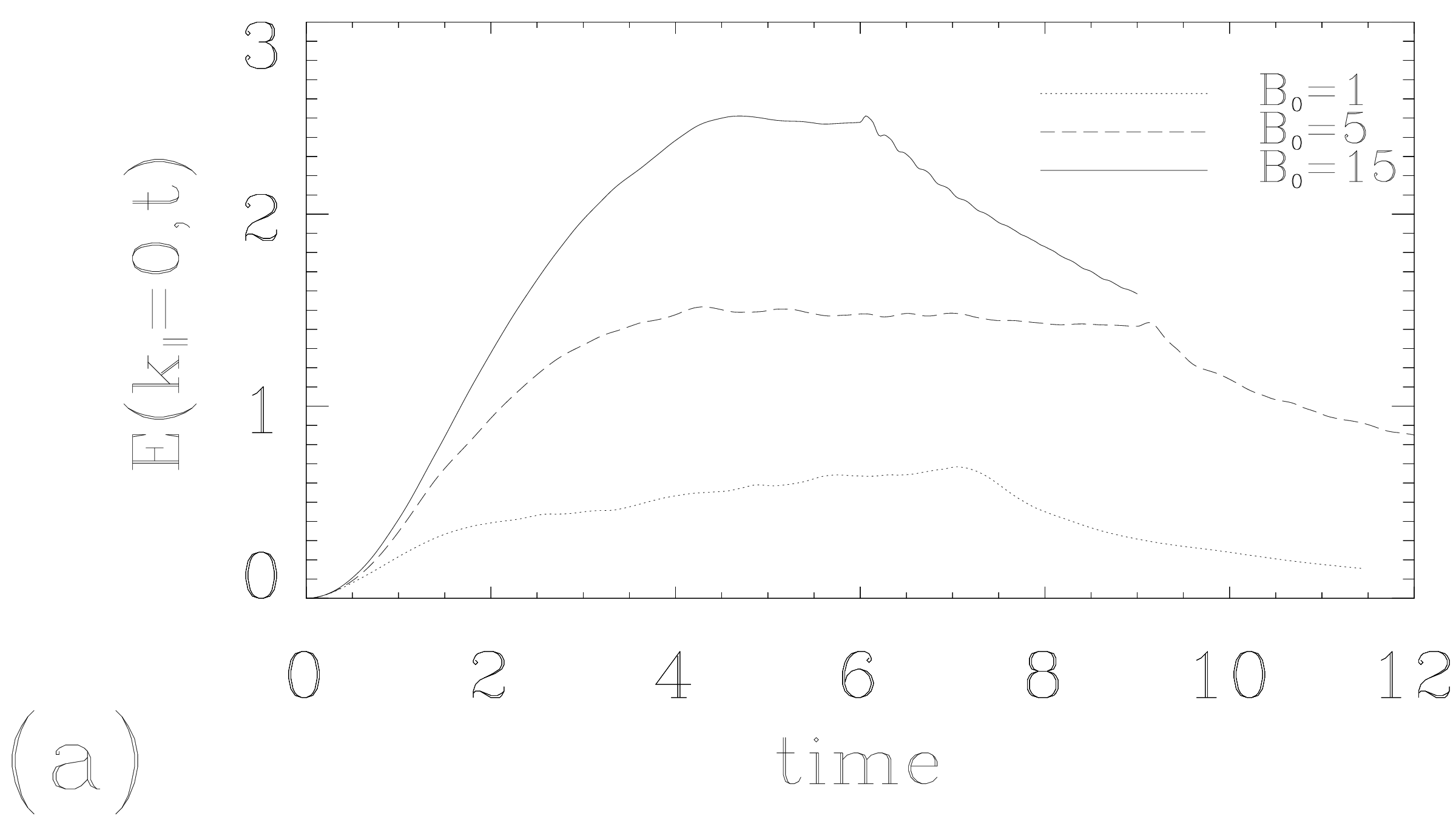}}
\resizebox{88mm}{!}{\includegraphics{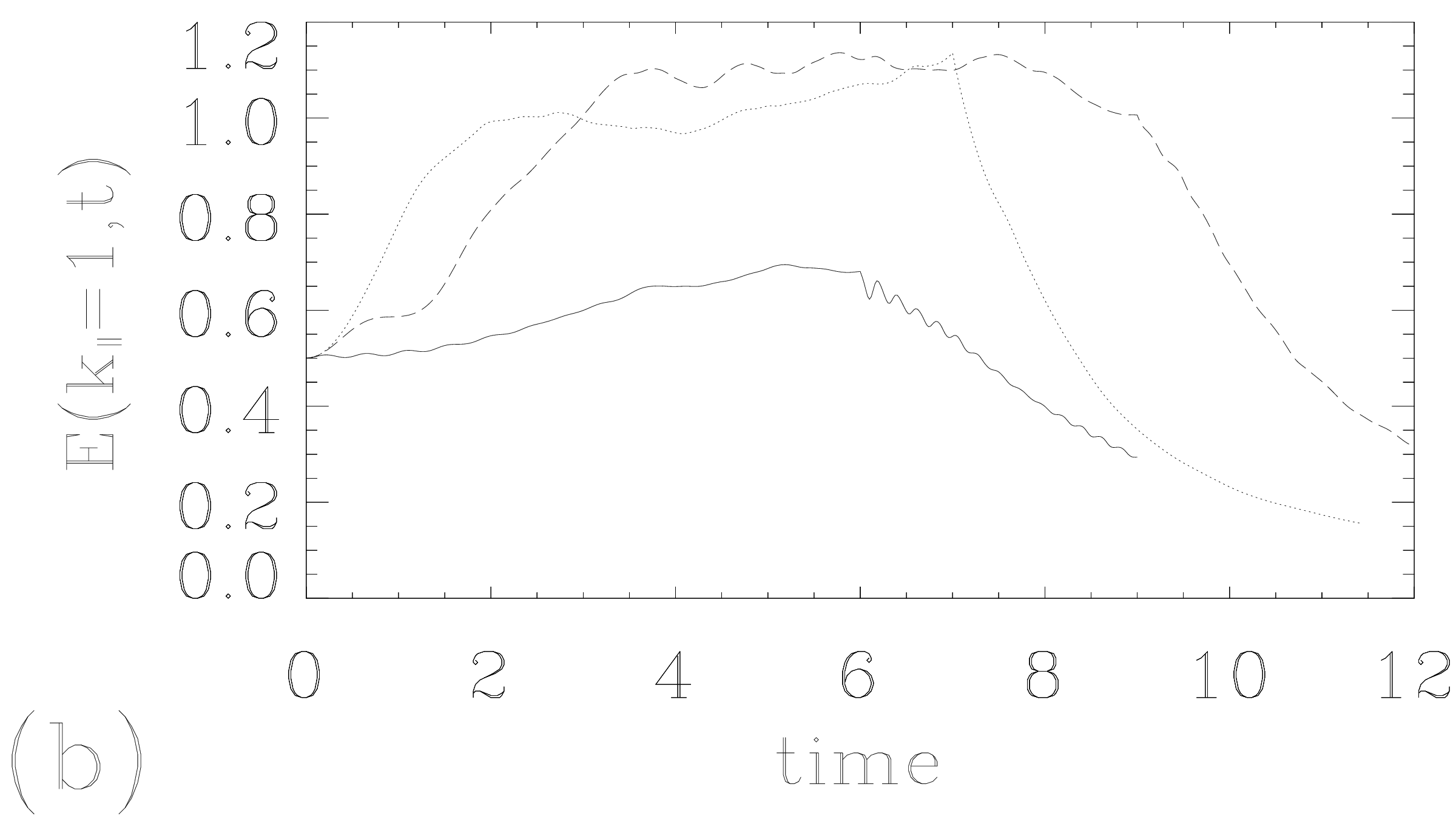}}\\
\resizebox{88mm}{!}{\includegraphics{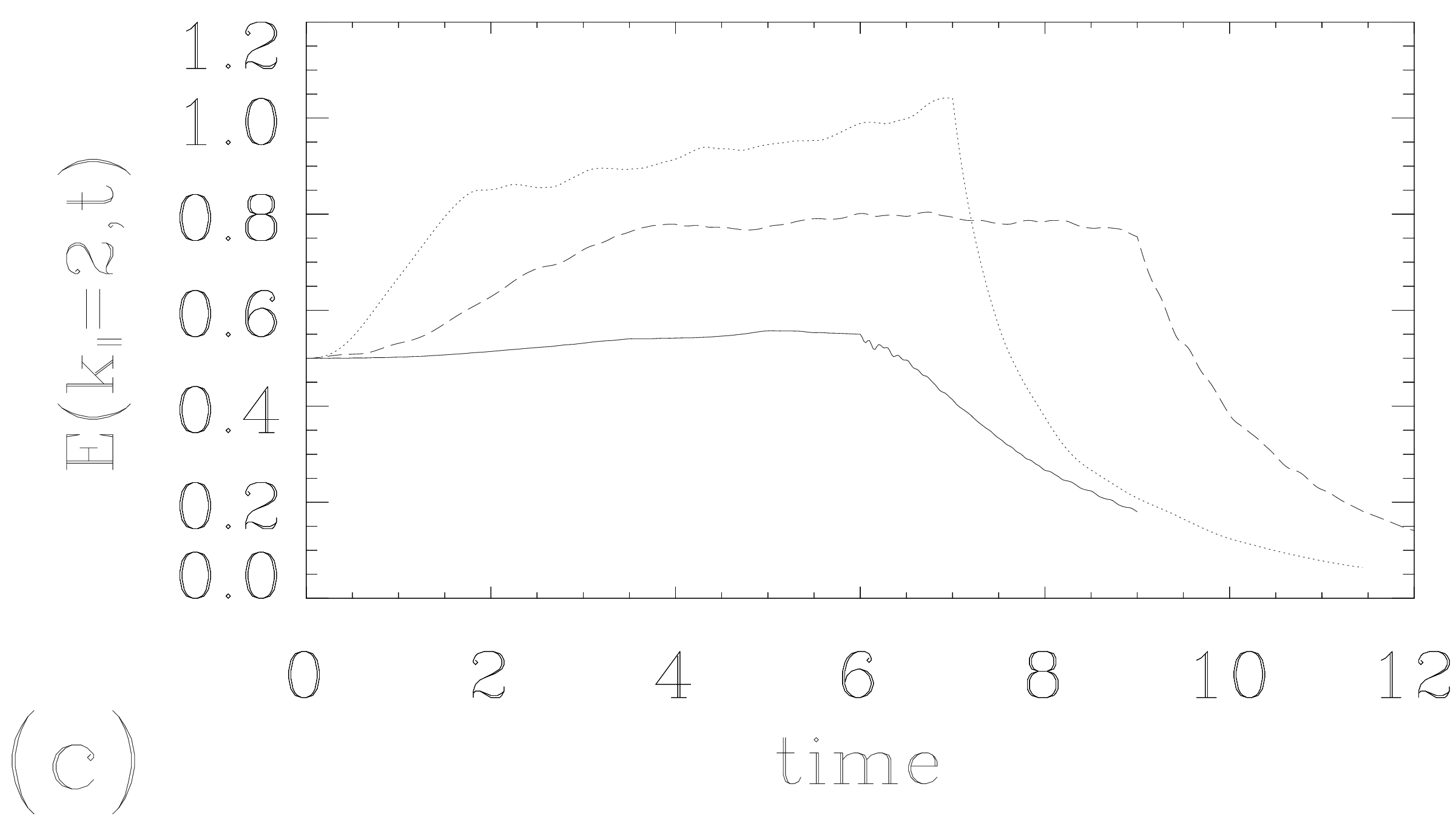}}
\resizebox{88mm}{!}{\includegraphics{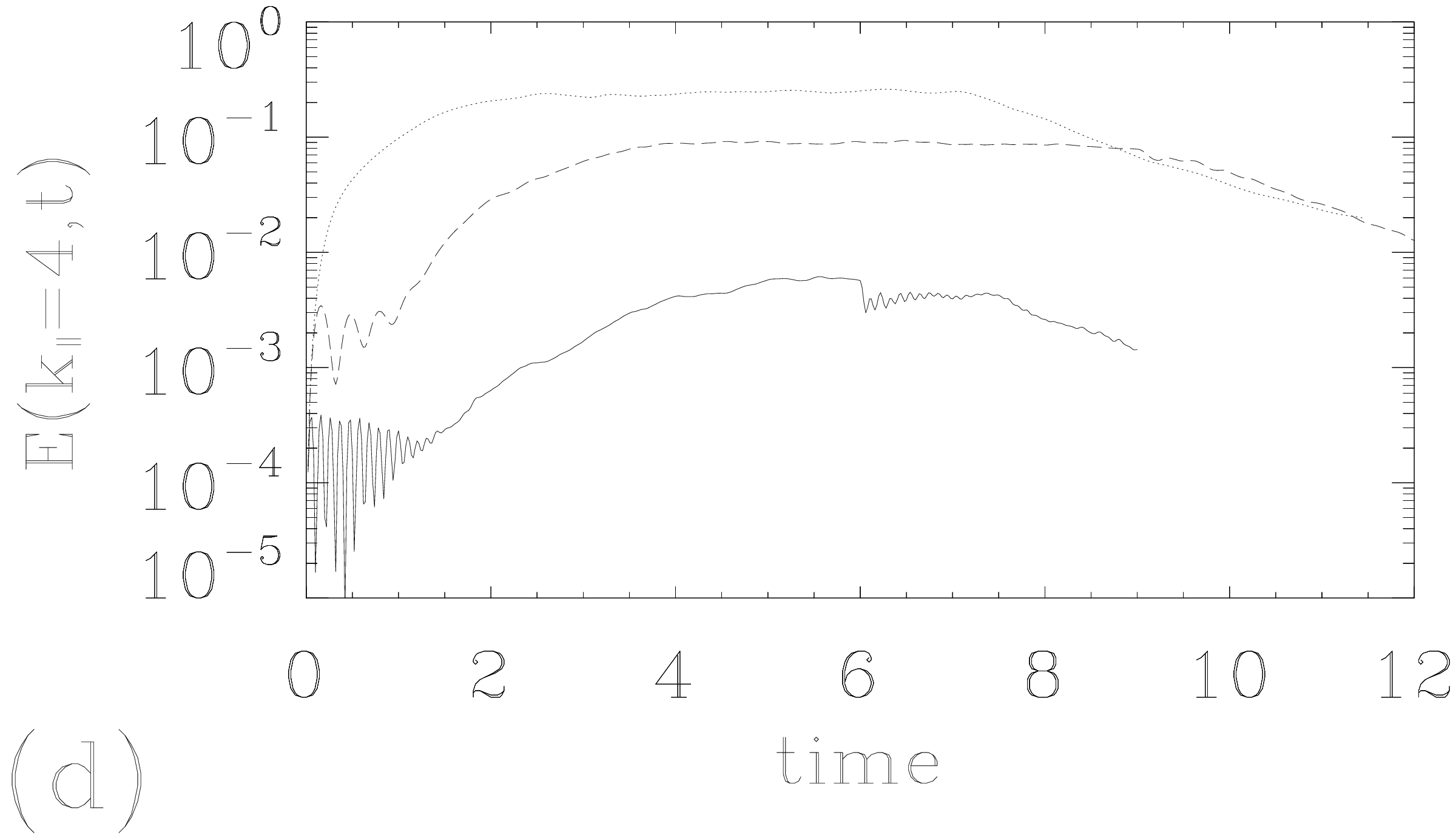}}
\caption{Temporal evolution of $E(\kpa,t)$ for $B_0=1$ (dot), $5$ (dash) and $15$ (solid) (runs ${\bf Ia}$ to 
${\bf IIIa}$) at $k_\pa=0$ (a), $1$ (b), $2$ (c) and $4$ (d). Note the use of semi-logarithmic coordinates for 
$k_\pa=4$ (bottom right). 
\label{FigtmpEkb0}}
\end{figure*}
Figure~\ref{FigtmpEkb0} shows the time variation of the total energy, namely
\be
E(\kpa,t) = E^+(\kpa,t) + E^-(\kpa,t) \, , 
\ee
with
\be 
E^\pm(\kpa,t) = \frac{1}{2} \int_{\kpe}{\bf \hat{z}}^{\pm^2}(\kpe,\kpa) d \kpe \, ,
\ee
where $\kpa$ is fixed to $0$, $1$, $2$ and $4$. We first note that the energy in plan $\kpa=1$, $2$ does not 
change very much at $B_0=15$ whereas it does change for other values of $B_0$. This behavior is 
surprising since the external force acts in particular on plans $\kpa=1$ and $2$. This situation contrasts 
with the energy in the $\kpa=0$ plan where the total energy exhibits a strong increase for $B_0=15$. 
For the plan $\kpa=4$ we note that the energy reaches only a small value when $B_0=15$. This observation 
demonstrates that the 2D state pumps efficiently the energy injected into the system at small parallel 
wavenumbers when a strong external magnetic field is applied. 
The second interesting observation in Fig.~\ref{FigtmpEkb0} is the oscillations found in the plan $\kpa=4$ 
(and also for higher values of $\kpa$; not shown) 
at the very beginning of runs ${\bf IIa}$ and ${\bf IIIa}$. During this period of time which extends 
approximately up to $t=1.5$ for $B_0=15$ and $t=1$ for $B_0=5$, the ratio $B_0/b_{rms}$ is larger than 
$10$ and $4$ for respectively $B_0=15$ and $5$. 
These oscillations are apparently correlated to the intensity of the mean magnetic field with a higher 
frequency at stronger $B_0/b_{rms}$. 
The last remark is about the dynamics at time $t>t^*$. For all simulations and all energies we note generally 
a decrease in time -- sometimes sharp like in $\kpa=1$, $2$ plans for $B_0=1$ -- which corresponds to a 
decay phase. Note that for $B_0=15$ the decay is characterized initially by small oscillations which are in 
phase opposition between the energies in plan $\kpa=0$ and $1$. This feature illustrates the energy exchange 
between the two first $\kpa$ plans. 

\begin{figure}[ht]
\resizebox{86mm}{!}{\includegraphics{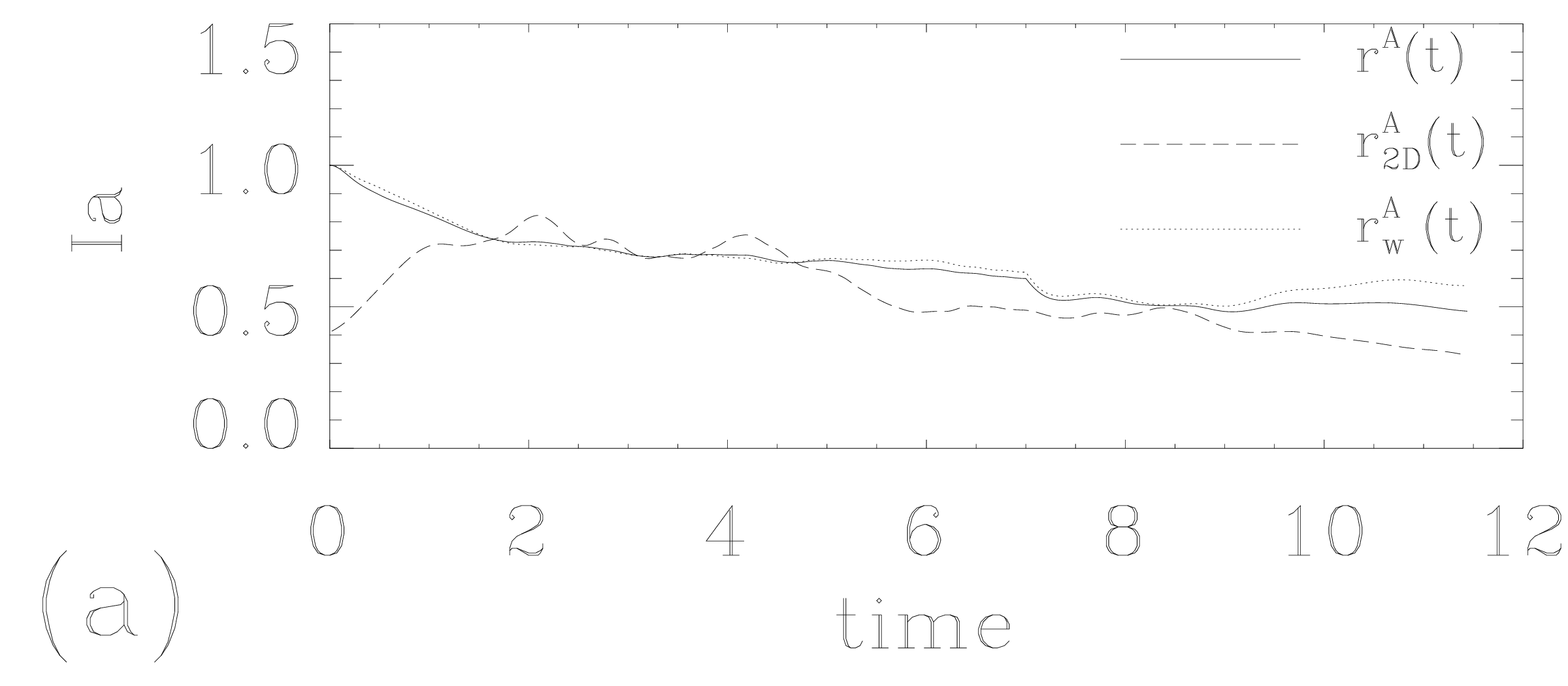}}\\
\resizebox{86mm}{!}{\includegraphics{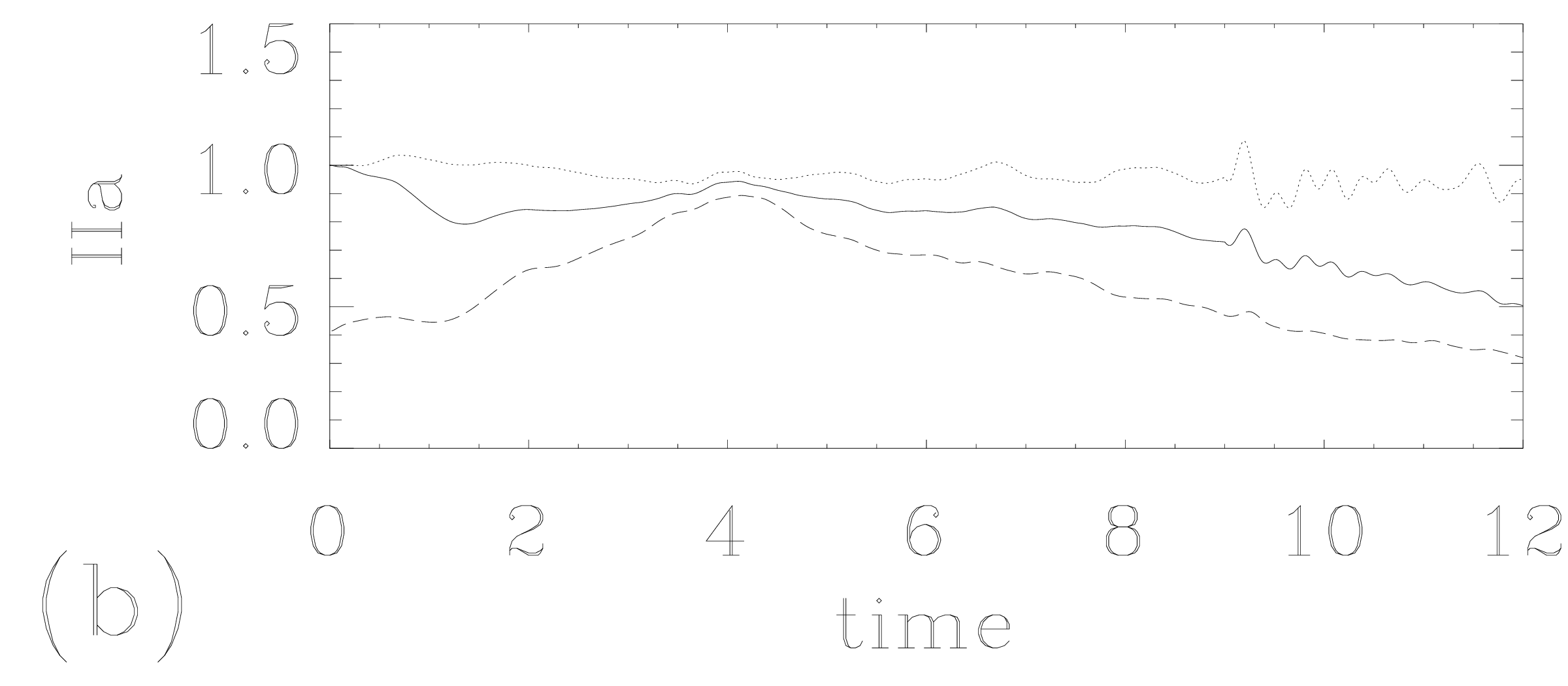}}\\
\resizebox{86mm}{!}{\includegraphics{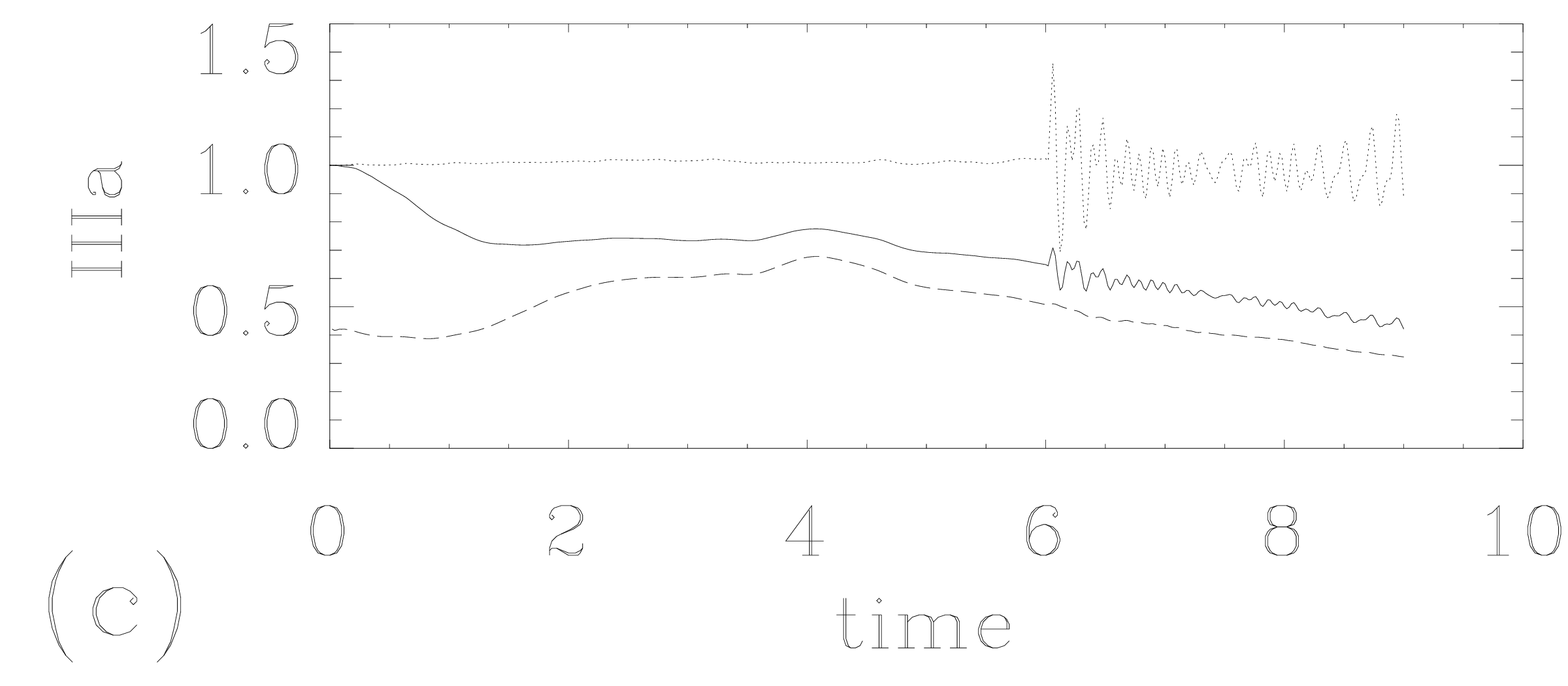}}
\caption{Temporal evolution of Alfv\'en ratios $r^A(t)$ (solid), $r^A_{2D}(t)$ (dash) and $r^A_w(t)$ (dot) for 
$B_0=1$ (a), $5$ (b) and $15$ (c) (runs ${\bf Ia}$ to ${\bf IIIa}$). 
\label{FigtmprapAkb0}}
\end{figure}
Figure~\ref{FigtmprapAkb0} presents the Alfv\'en ratios between  the kinetic and magnetic energies 
for the 3D modes 
\be
r^A_{w}(t)= {E_w^v(t) \over E_w^b(t)} \, ,
\label{eqrAw}
\ee 
the $2$D state 
\be
r^A_{2D}(t)= {E_{2D}^v(t) \over E_{2D}^b(t)} \, ,
\label{eqrA2D}
\ee 
and the total flow  
\be
r^A(t)= {E^v(t) \over E^b(t)} \, ,
\label{eqrA}
\ee 
where the definitions (\ref{deft}), (\ref{defw}) and (\ref{def2d}) have been directly applied to the velocity and 
magnetic fields. The Alfv\'en ratio allows us to measure the prevalence of Alfv\'en wave fluctuations: 
for example, in the wave turbulence regime one can demonstrate at the level of the kinematics 
an equipartition between the kinetic and magnetic energies \citep{Galtier2000}. Therefore, a departure 
from unity suggests the presence of non Alfv\'enic fluctuations \citep{Bigot08b}. 
In simulations ${\bf IIa}$ and ${\bf IIIa}$, the 3D modes are in equipartition with $r^A_{w} \sim 1$. However, 
we may note a difference between $t<t^*$ and $t>t^*$: a lack of oscillations is found during the driving phase 
whereas significant oscillations happen during the decay phase which are stronger for $B_0=15$. 
The 2D state evolves quiet differently with an Alfv\'en ratio $r^A_{2D}$ significantly smaller than the unity for 
runs ${\bf Ia}$ to ${\bf IIIa}$ which means that the 2D state is magnetically dominated. We also note that 
$r^A_{2D}$ seems not to be strongly affected by the value of the mean field $B_0$ with first an increase 
(from the initial value at $0.4$) and then a slight decrease. The behavior of the Alfv\'en ratio for the 2D state 
is therefore quiet similar to what was found in the pure decay regime \citep{Bigot08b}. 
The time variation of $r^A$ is smooth with initially only a slight decrease after which the value stabilizes 
around $0.8$ for $B_0=5$ and $15$. When the driving is suppressed a slight decrease with small oscillations 
is found as expected since it is the addition of the effects of $r^A_{w}$ and $r^A_{2D}$. 
For run {\bf Ia} where $B_0=1$ a different behavior is observed since all Alfv\'en ratios are less than unity 
and no Afv\'enic fluctuations are detected. This behavior is partly explained by the increase of $b_{rms}$ 
which becomes larger than $B_0$ and prevent any dynamical effect of the mean magnetic field on the flow. 

\begin{figure}[ht]
\resizebox{86mm}{!}{\includegraphics{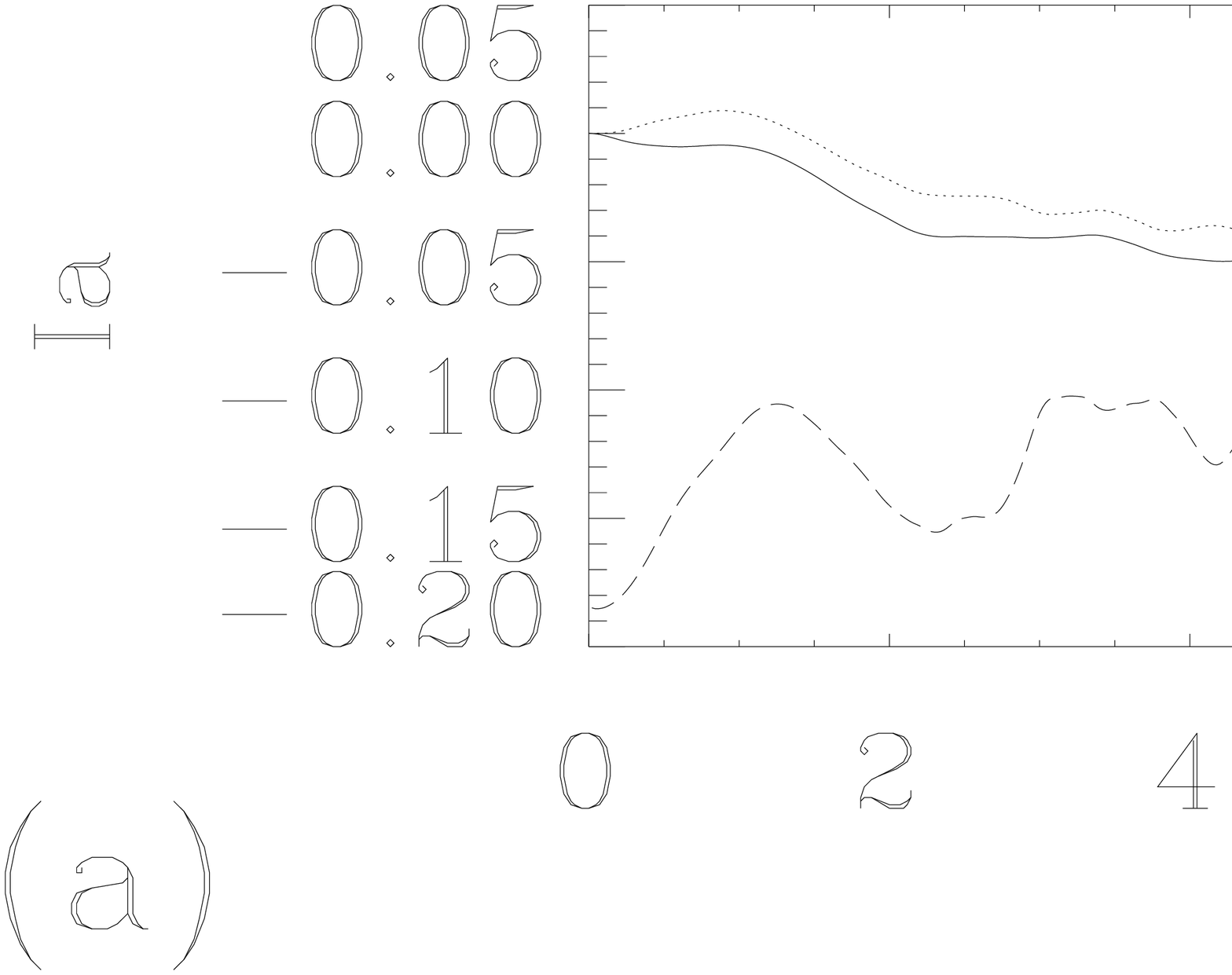}}
\resizebox{86mm}{!}{\includegraphics{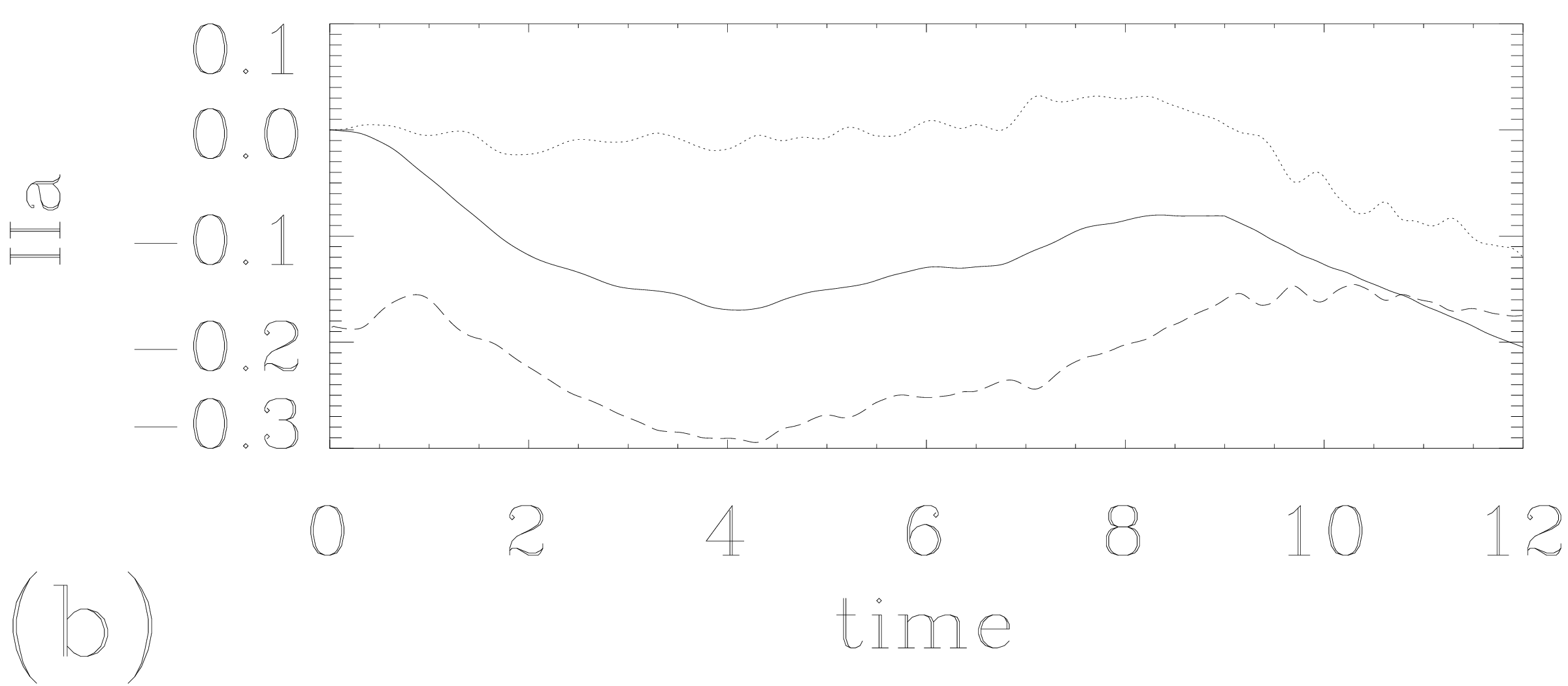}}
\resizebox{86mm}{!}{\includegraphics{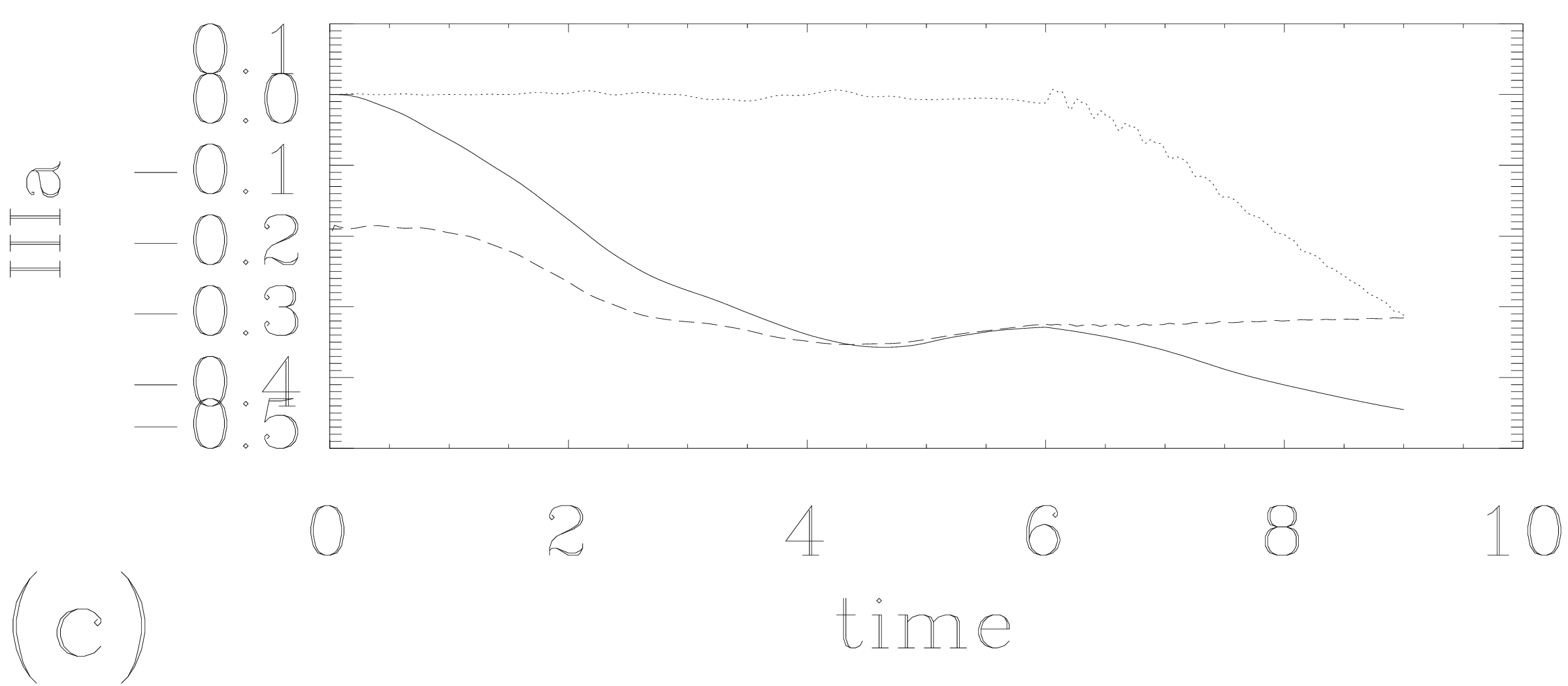}}
\caption{Temporal evolution of $\rho(t)$ (solid), $\rho_{2D}(t)$ (dash) and $\rho_w(t)$ (dot) for $B_0=1$ (a), 
$5$ (b) and $15$ (c) (runs ${\bf Ia}$ to ${\bf IIIa}$). 
\label{Figtmpcorkb0}}
\end{figure}
Figure~\ref{Figtmpcorkb0} shows the reduced cross-helicity $\rho($t$)$ between the velocity and magnetic 
fields which was already defined by relation (\ref{rho}). In addition, we may define the reduced cross-helicity 
for the 3D modes and the 2D state as respectively 
\be
\rho_w (t)= {E_w^+(t) - E_w^-(t) \over E_w^+(t) + E_w^-(t)} \, ,
\label{rho1}
\ee
and
\be
\rho_{2D}(t)= {E_{2D}^+(t) - E_{2D}^-(t) \over E_{2D}^+(t) + E_{2D}^-(t)} \, ,
\label{rho2}
\ee
where
\be 
E_{2D}^\pm(t) = \frac{1}{2} \int_{k_\perp}{\bf \hat{z}}^{\pm^2}(k_\perp,k_\pa=0) dk_\perp \, .
\ee
We remind that the cross-helicity is a measure of the relative amount of Alfv\'en wave packets which 
propagate in opposite directions along the uniform magnetic field ${\bf B_0}$. It is generally found that 
MHD turbulence evolves towards a state of maximal cross-helicity with an alignment or an anti-alignment 
of ${\bf v}$ and ${\bf b}$ according to the sign of the cross-helicity \citep{Grappin83}. 
This definition applies to (\ref{rho1}) which is a refined definition of the cross-helicity where the non 
propagating part is discarded. With the definition (\ref{rho2}) the situation is different. If we assume 
isotropy in the $\kpa=0$ plan then $\rho_{2D}$ can be seen as a local measure of the cross-helicity 
since we can always define a local mean magnetic field ${\bf b_0}$ along which 2D Alfv\'en fluctuations
propagate. In Fig.~\ref{Figtmpcorkb0} we see that the reduced cross-helicity of the 3D modes is initially null 
and slightly evolves during the driving phase for $B_0=1$; for $B_0=5$ only small oscillations are detected 
whereas for $B_0=15$ it is almost constant. One has to wait the decay phase to see a significant 
modification of $\rho_w$. 

The situation is different for the reduced cross-helicity of the 2D state which experiences a variation whatever 
the intensity of the mean field $B_0$ is. Note that this variation is moderate since it never exceeds in 
absolute value $0.35$. Note also that our initial condition leads to a negative value for $\rho_{2D}$ 
since we impose a condition only on the global cross-helicity, \ie $\rho(t=0) = 0$ ($\rho_{2D}$ cannot be 
defined initially since the 2D state has no energy). 
In other words, we generate initially more $z_{2D}^-$ than $z_{2D}^+$ fluctuations. For $B_0=5$ and 
$15$ the time at which the stationary phase is reached are $t \sim 4$ and $5$ respectively; at these 
times, $\rho_{2D}$ indicates around $30 \%$ more $z_{2D}^-$ than $z_{2D}^+$. Clearly, we see that 
the evolution of $\rho(t)$ is mainly affected by the evolution of $\rho_{2D}(t)$. Therefore, it seems to 
be important in this problem to make the distinction between the cross-helicity of the 3D modes and 
the 2D state to evaluate if whether or not we have the domination of one type of Alfv\'en wave packet
propagating along the external magnetic field ${\bf B_0}$. Another interesting behavior is the variation 
of $\rho_{2D}$ during the decay phase: we observe a weaker variation for $B_0=5$ and $15$ with in the 
latter case an almost constant reduced cross-helicity.

\section{Spectral analysis}
\label{sec5}

In this Section, we study the spectral behavior of driven MHD turbulence. We start from the 
bidimensional Els\"asser energy spectra, $E^\pm(k_\perp,k_\pa)$, and define respectively the 
corresponding unidimensional spectra for the 3D modes 
\be
E_w^\pm(k_\perp) = \int_{k_\pa>0} E^\pm(k_\perp,k_\pa) dk_\pa \, ,
\label{specwaves}
\ee
and the $2$D state 
\be
E_{2D}^\pm(k_\perp) = E^\pm(k_\perp,k_\pa=0)  \, .
\label{spec2D}
\ee
Then, the addition of both defines the usual unidimensional Els\"asser energy spectrum
\be
E^\pm(k_\perp) = \int_{k_\pa\geq0} E^\pm(k_\perp,k_\pa) dk_\pa \, .
\label{spec}
\ee
The comparison will be made between all runs, \ie runs ${\bf Ia}$ to ${\bf IIIa}$ and run ${\bf IIIb}$.  

\begin{figure*}[ht]
\resizebox{170mm}{!}{\includegraphics{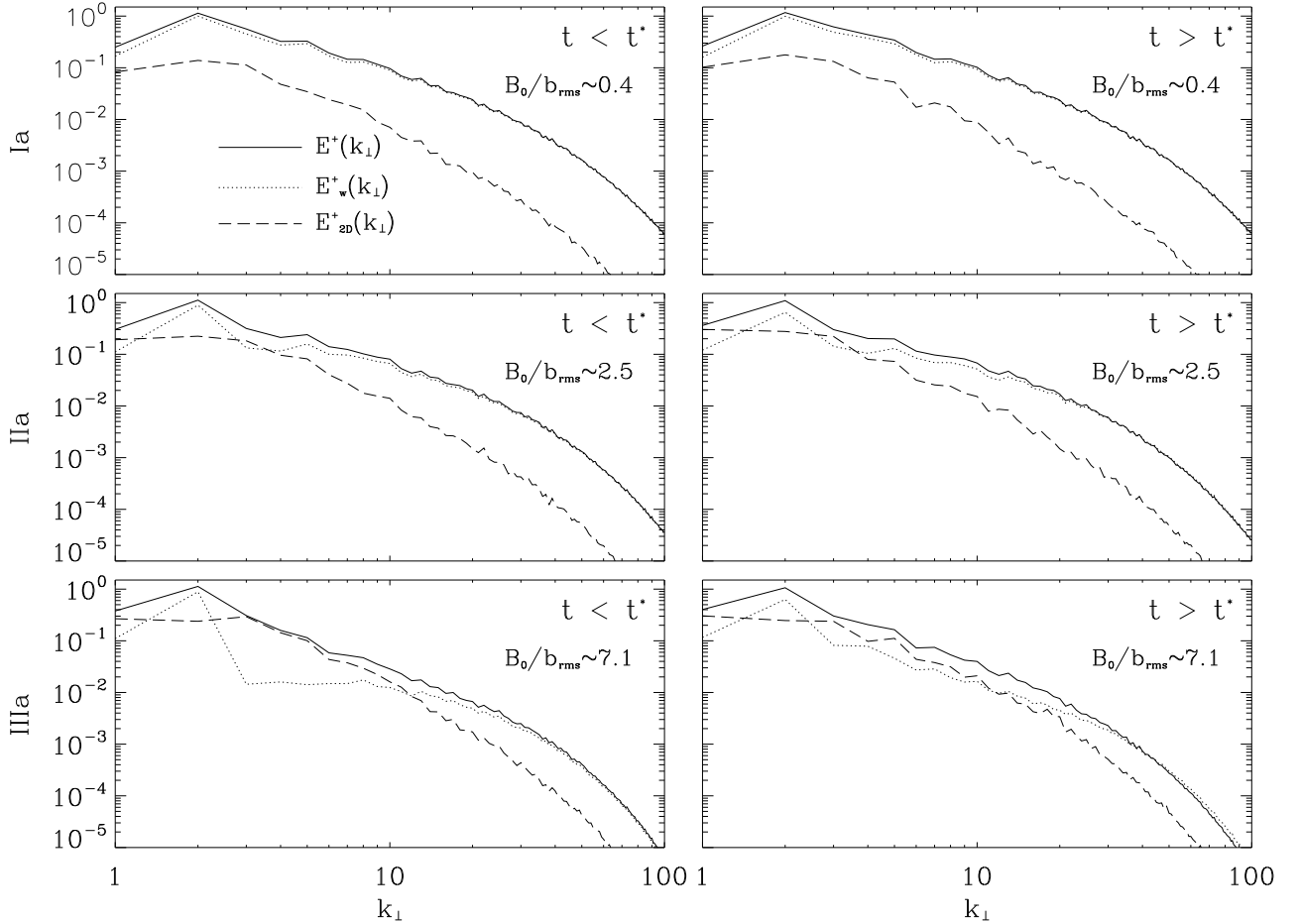}}
\caption{Energy spectra $E^+(\kpe)$ (solid), $E^+_{2D}(\kpe)$ (dash) and $E^+_w(k_\perp)$ (dot) for 
$B_0=1$ (top), $B_0=5$ (middle) and $B_0=15$ (bottom) which corresponds to runs ${\bf Ia}$, ${\bf IIa}$ 
and ${\bf IIIa}$ respectively. Spectra are displayed at times $t<t^*$ (left) and $t>t^*$ (right) which 
corresponds to $t=5$ and $7.2$ for $B_0=1$, $t=5.1$ and $9.2$ for $B_0=5$, $t=5.8$ and $6.2$ for $B_0=15$. 
\label{Fig1_specep}}
\end{figure*}

\subsection{Runs Ia to IIIa}

Figure~\ref{Fig1_specep} displays energy spectra for the Els\"asser field $E^+(\kpe)$, the 3D modes 
$E_w^+(\kpe)$ and the $2$D state $E_{2D}^+(\kpe)$, at different $B_0$ intensity and with the same initial 
condition (runs {\bf Ia} to {\bf IIIa}).  The comparison is made between two times, $t<t^*$ and $t>t^*$, where 
$t^*$ is the time from which the external force is suppressed. 
Shortly after the beginning of the simulations we observe that the energy spectrum $E^+_{2D}(\kpe)$ loads 
up. For $B_0=1$ this spectrum remains for all times about one order of magnitude smaller than $E_w^+(\kpe)$. 
The situation is different for stronger magnetic intensity like $B_0=15$. In this case, we see that 
$E^+_{2D}(\kpe)$ may dominate $E_w^+(k_\perp)$ at small perpendicular wavenumbers, \ie $\kpe<10$. 
Then, the spectrum $E^+(\kpe)$ is formed at large scales mainly by the 2D state and at small scales by the 
3D modes. Run {\bf IIa} is an intermediate case where the 2D state becomes dominant in a very narrow 
range of perpendicular wavenumbers. 
The second interesting comment is about the break observed in some 3D modes spectrum just after the 
domain of excitation which is limited in particular to $\kpe \in [1,2]$ (see relation (\ref{f1})). This break appears 
for strong $B_0$ and during the driving phase. Indeed, shortly after time $t^*$ an extended power law energy 
spectrum is formed. At this resolution the driving prevents the possibility to measure any power law for the 
energy spectrum of 
the 3D modes; it is believed that it could also artificially modify the power law of the Els\"asser energy spectra. 
Note that a similar break was already observed in driven MHD turbulence \citep{MBG03}. 
Finally, at time larger than $t^*$ the system evolves freely and the slope at the lowest $\kpe$ is modified slightly 
for strong $B_0$ and strongly for $B_0=1$ with a flattening for the energy spectrum of the 3D modes: the slope
goes from $\kpe^3$ initially to roughly $\kpe^2$ after $t^*$. 
Note that $E^-(\kpe)$, $E^-_w(\kpe)$ and $E^-_{2D}(\kpe)$ (not shown) exhibit a similar behavior. 

\begin{figure*}[ht]
\resizebox{170mm}{!}{\includegraphics{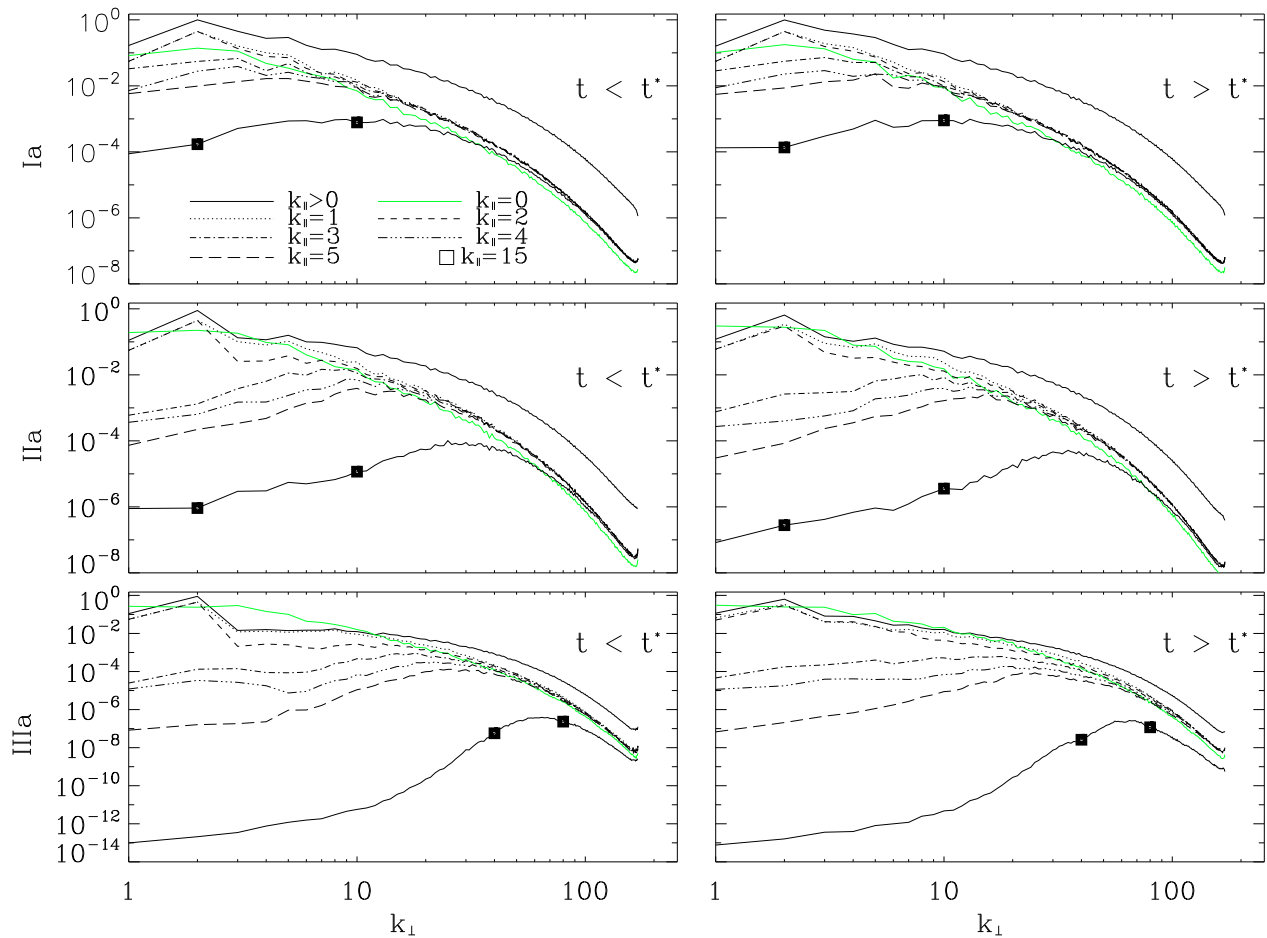}}
\caption{Energy spectra $E^+(\kpe,\kpa)$ at fixed $\kpa$ and for $B_0=1$ (top), $B_0=5$ (middle) and 
$B_0=15$ (bottom) which correspond to runs ${\bf Ia}$, ${\bf IIa}$ and ${\bf IIIa}$ respectively. Spectra 
are displayed at times $t<t^*$ (left) and $t>t^*$ (right) which are the same as in Fig.~\ref{Fig1_specep}. 
The values chosen for $\kpa$ are $1,2,3,4,5,15$. For comparison we also plot the spectra of the 2D 
state (green color online, grey line) for which $\kpa=0$ and of the 3D modes. 
\label{Fig2_specp}}
\end{figure*}
The bidimensional Els\"asser energy spectra $E^+(\kpe,\kpa)$ are plotted in Fig.~\ref{Fig2_specp} at 
fixed $\kpa$. We choose the values $\kpa=0$ to $5$ and also $\kpa=15$. As in Fig.~\ref{Fig1_specep} 
a comparison is made between two times, $t<t^*$ and $t>t^*$. The most remarkable feature is that the 
spectra at fixed $\kpa$ are characterized by a pinning effect with a convergence of the spectra at large 
$\kpe$. The case $B_0=15$ is an exception where the pinning is observed only for close values of $\kpa$ 
as we can see with a significant difference for $\kpa=15$. Note that the spectrum of the 2D state has a 
slightly different behavior since it does not converge towards the other spectra at large $\kpe$. 
The second observation is that at times larger than $t^*$ power laws are detected for spectra with $\kpa=0$ 
to $5$, with a reduction of the inertial range while the power law index seems unchanged. The inertial 
ranges are then shifted to larger $\kpe$. 

\begin{figure*}[ht]
\resizebox{170mm}{!}{\includegraphics{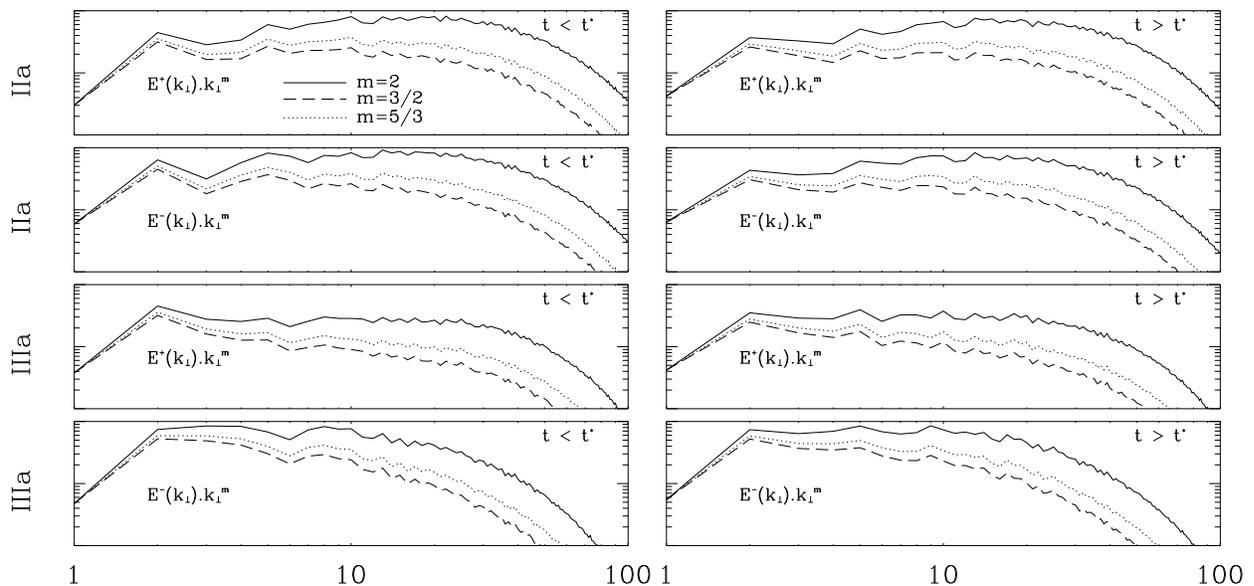}}
\caption{Energy spectra $E^\pm(\kpe)$ compensated by $\kpe^2$ (solid), $\kpe^{3/2}$ (dash) and 
$\kpe^{5/3}$ (dot) at times $t<t^*$ (left), $t>t^*$ (right), for runs ${\bf IIa}$ and ${\bf IIIa}$. Note that the 
times are the same as in Fig.~\ref{Fig1_specep}. 
\label{Fig3_spec}}
\end{figure*}
Figure~\ref{Fig3_spec}  presents unidimensional Els\"asser energy spectra $E^\pm (\kpe)$ defined by 
relation (\ref{spec}). These spectra are compensated by different power laws $\kpe^m$ with $m=3/2$, $5/3$ 
and $2$ which correspond to the predictions for strong and weak turbulence. Only runs {\bf IIa} and {\bf IIIa} 
are shown for which respectively $B_0/b_{rms} = 2.5$ and $7.1$ (run {\bf Ia} behaves similarly to run {\bf IIa}). 
We see that run {\bf IIa} fits well with power laws like $\kpe^{-3/2}$ or $\kpe^{-5/3}$ whereas run {\bf IIIa} is 
close to $\kpe^{-2}$. These results are in relatively good agreement with the characteristic timescales given 
in Table~\ref{table1} which satisfy relations $\tau_A \sim \tau_{nl}$ for run {\bf IIa} and $\tau_{A} \ll \tau_{nl}$ 
for run {\bf IIIa}. We also see that for run {\bf IIIa} the inertial range is significantly larger for $E^+$ than $E^-$. 
We will come back to this point below and show that the 2D state may explain such a feature.

\subsection{Runs IIIa and IIIb}

\begin{figure*}[ht]
\resizebox{170mm}{!}{\includegraphics{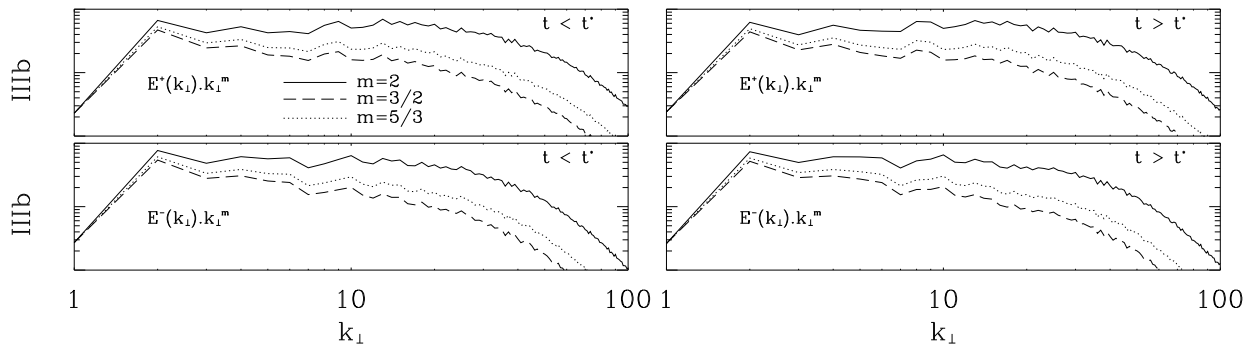}}
\caption{Energy spectra $E^\pm(\kpe)$ (run ${\bf IIIb}$) for times $t=t^*=5.1$ (left) and $t=5.2$ (right) 
compensated by $\kpe^2$ (solid), $\kpe^{3/2}$ (dash) and $\kpe^{5/3}$ (dot). 
\label{Fig3_speci1}}
\end{figure*}
In this Section a comparison is made between runs {\bf IIIa} and {\bf IIIb} for which $B_0=15$. The difference 
between these runs comes from the initial conditions and the external force (see Table~\ref{table1} and 
Sec.~\ref{IC}). 

Figure~\ref{Fig3_speci1} shows the compensated energy spectra $E^\pm(\kpe)$ for run $\bf{IIIb}$. As in 
Fig. \ref{Fig3_spec} spectra are compensated by different power laws $\kpe^m$ with $m=3/2$, $5/3$ and 
$2$. We observe the same characteristics 
as for run $\bf{IIIa}$ with in particular timescales compatible with relation $\tau_{A} \ll \tau_{nl}$ (see 
Table~\ref{table1}). The only one difference is the size of the inertial ranges which are about the same for 
both spectra. 

\begin{figure}[ht]
\resizebox{86mm}{!}{\includegraphics{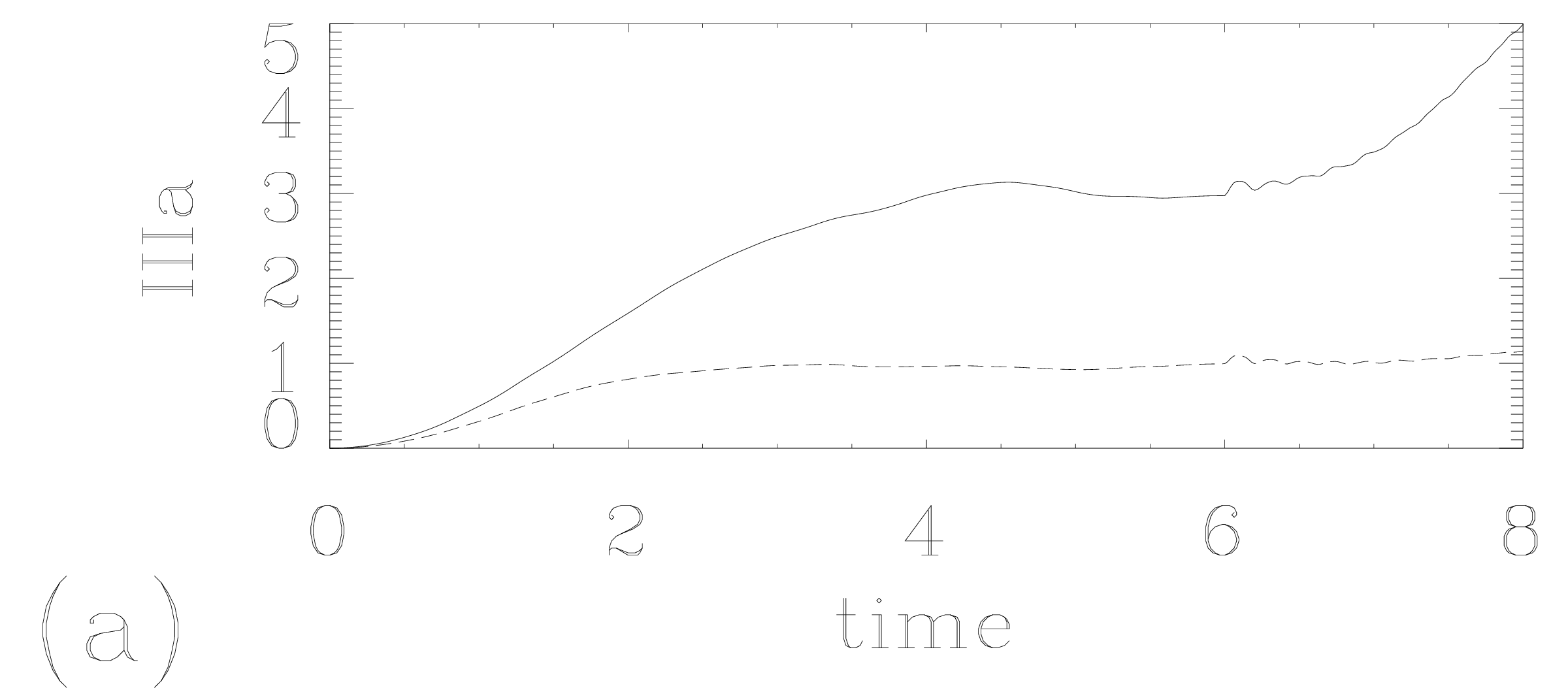}}
\resizebox{86mm}{!}{\includegraphics{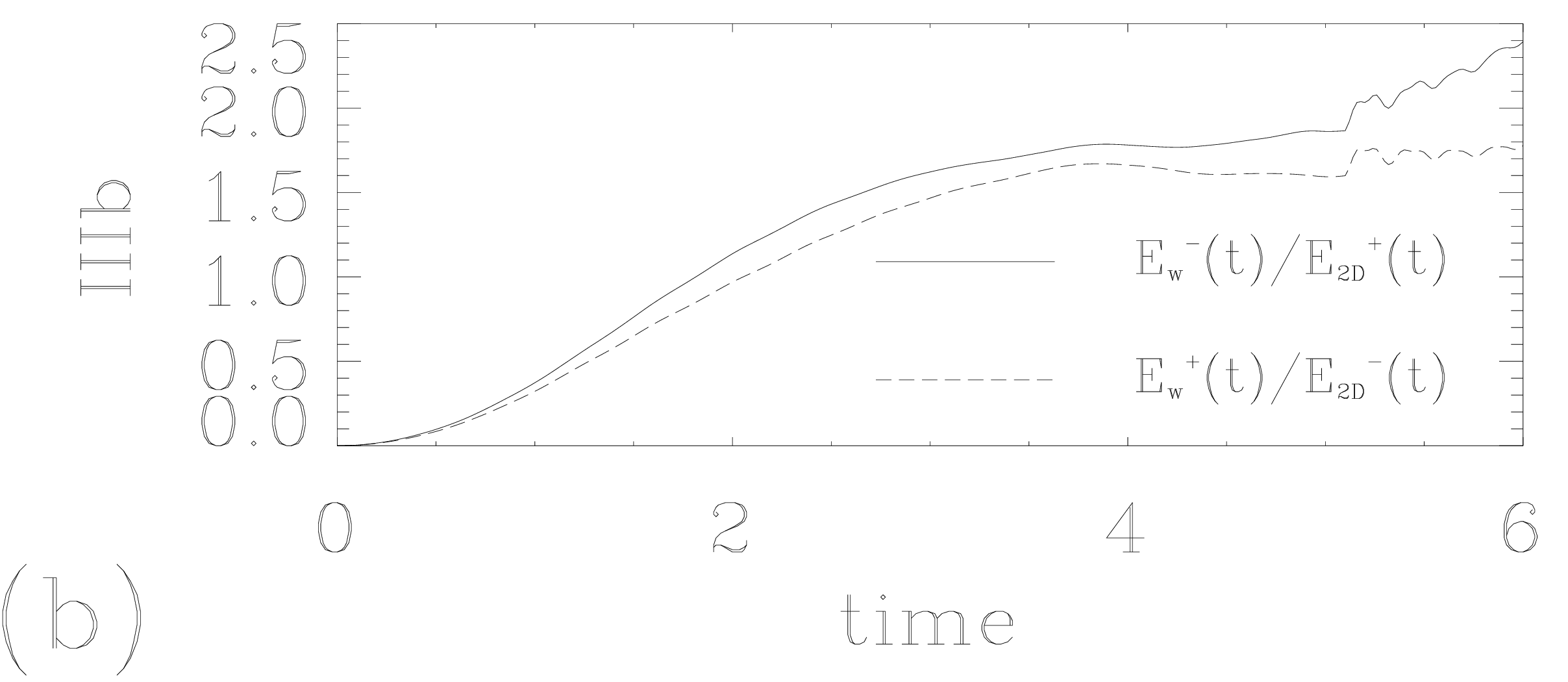}}
\caption{Temporal evolution of the ratios $E_w^\mp/E_{2D}^\pm$ for runs ${\bf IIIa}$ (top) and ${\bf IIIb}$ (bottom).
\label{Fig_ratiowturb}}
\end{figure}
In order to explain the previous observations and the difference between cases $\bf{IIIa}$ and $\bf{IIIb}$ we shall 
analyze the ratios $E_w^{\mp}/E_{2D}^{\pm}$. We first remind that in the regime of wave turbulence the 
wavevectors (${\bf k}$, ${\bf p}$, ${\bf q}$) satisfy the resonant triadic interactions ${\bf k} = {\bf p} + {\bf q}$ 
and for example $\omega({\bf k}) = \omega({\bf p}) - \omega({\bf q})$ which leads to the resonance condition 
$q_\pa=0$ and to the prediction that only a spectral transfer transverse to ${\bf B_0}$ happens. It also 
means that we always have an interaction between an Alfv\'en wave packet of one type of polarity and the 
fluctuations of the 2D state with the opposite polarity. Numerical simulations have confirmed that the resonant 
interactions become dominant for large $B_0/b_{rms}$ \cite{Alexakis07b}. 
Therefore, the amount of energy in the 2D state is an important parameter for the nonlinear dynamics. For example, 
the absence of energy inside the 2D state cannot lead to the development of a wave turbulence regime. 
In Fig.~\ref{Fig_ratiowturb} the ratios $E_w^{\mp}/E_{2D}^{\pm}$ between the energies of the 3D modes and the 
2D state with opposite polarities are given. For run ${\bf IIIb}$, we see that these ratios are almost the same and 
reach roughly a stationary state for times $t \in [4,6]$ with values close to $1.5$. 
In this regime we have  $E_{2D}^+ \sim 0.59 E_w^-$ 
and $E_{2D}^- \sim 0.66 E_w^+$ which lead to similar spectra $E^\pm(\kpe)$ (see Fig.~\ref{Fig3_speci1}) with 
inertial ranges of the same size. For run ${\bf IIIa}$ the ratios are significantly different: in the stationary phase for
which $t \in [4,6]$ we have $E_{2D}^+ \sim 0.33 E_w^-$ and $E_{2D}^- \sim E_w^+$. In practice, the energy 
in the 2D state with a positive polarity is reduced which leads to a weakening of the nonlinear dynamics 
and a reduction of the inertial range for $E_w^-(\kpe)$ as we can see in Fig.~\ref{Fig3_spec}. Note that similarly
to run ${\bf IIIa}$ (see Fig~\ref{FigtmpEb0}) the energy contained in the 2D state for run ${\bf IIIb}$ represents 
$2/3$ of the total energy. 
\begin{figure}[ht]
\resizebox{86mm}{!}{\includegraphics{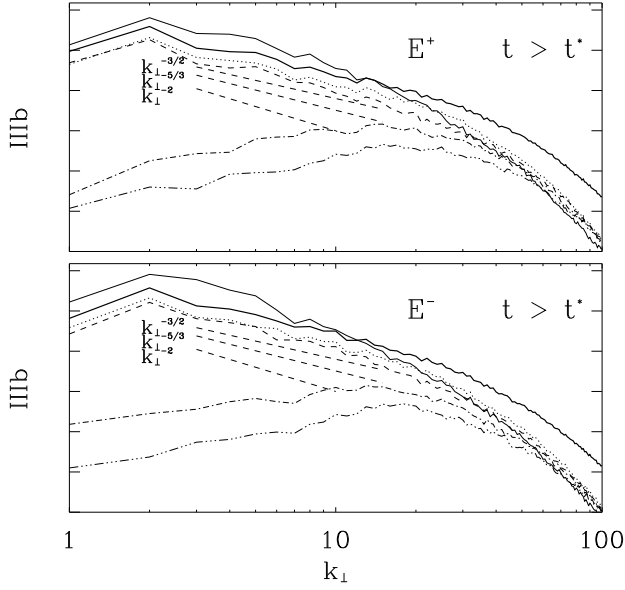}}
\caption{Energy spectra $E^+(\kpe,\kpa)$ (top) and $E^-(\kpe,\kpa)$ (bottom) at fixed $\kpa$ (run ${\bf IIIb}$) and
at time $t=5.5>t^*$. The values chosen for $\kpa$ are $1,2,3,4$ (same notation as in Fig. \ref{Fig2_specp}). 
For comparison we also plot the spectra of the 2D state (solid) and of the 3D modes (bold solid). 
\label{Fig2_spepmci1}}
\end{figure}

Figure~\ref{Fig2_spepmci1} displays spectra $E_w^\pm(\kpe)$, $E_{2D}^\pm(\kpe)$ and $E^\pm(\kpe,\kpa)$ for 
$k_\pa=1,2,3,4$. Globally we observe the same features as for run ${\bf IIIa}$ with a stronger domination of the 2D 
state at large scales. Note that at the smallest $\kpe$, the initial power law in $\kpe^2$ applied to spectra $\kpa=1,2$ 
is slightly modified and the scaling for $E_w^\pm(\kpe)$ is unchanged. Note also that the break previously 
reported for times $t<t^*$ is also observed (not shown). 
Finally, as we have already seen in Fig~\ref{Fig1_specep} (run $\bf{IIIa}$), for times $t>t^*$ an inertial range 
appears for spectra $E_w^\pm(\kpe)$ and  $E^\pm(\kpe,\kpa)$ with $k_\pa=1,2,3,4$ which seem roughly better 
fitted by $\kpe^{-3/2}$ or $\kpe^{-5/3}$ than $\kpe^{-2}$.

\begin{figure}[ht]
\resizebox{88mm}{!}{\includegraphics{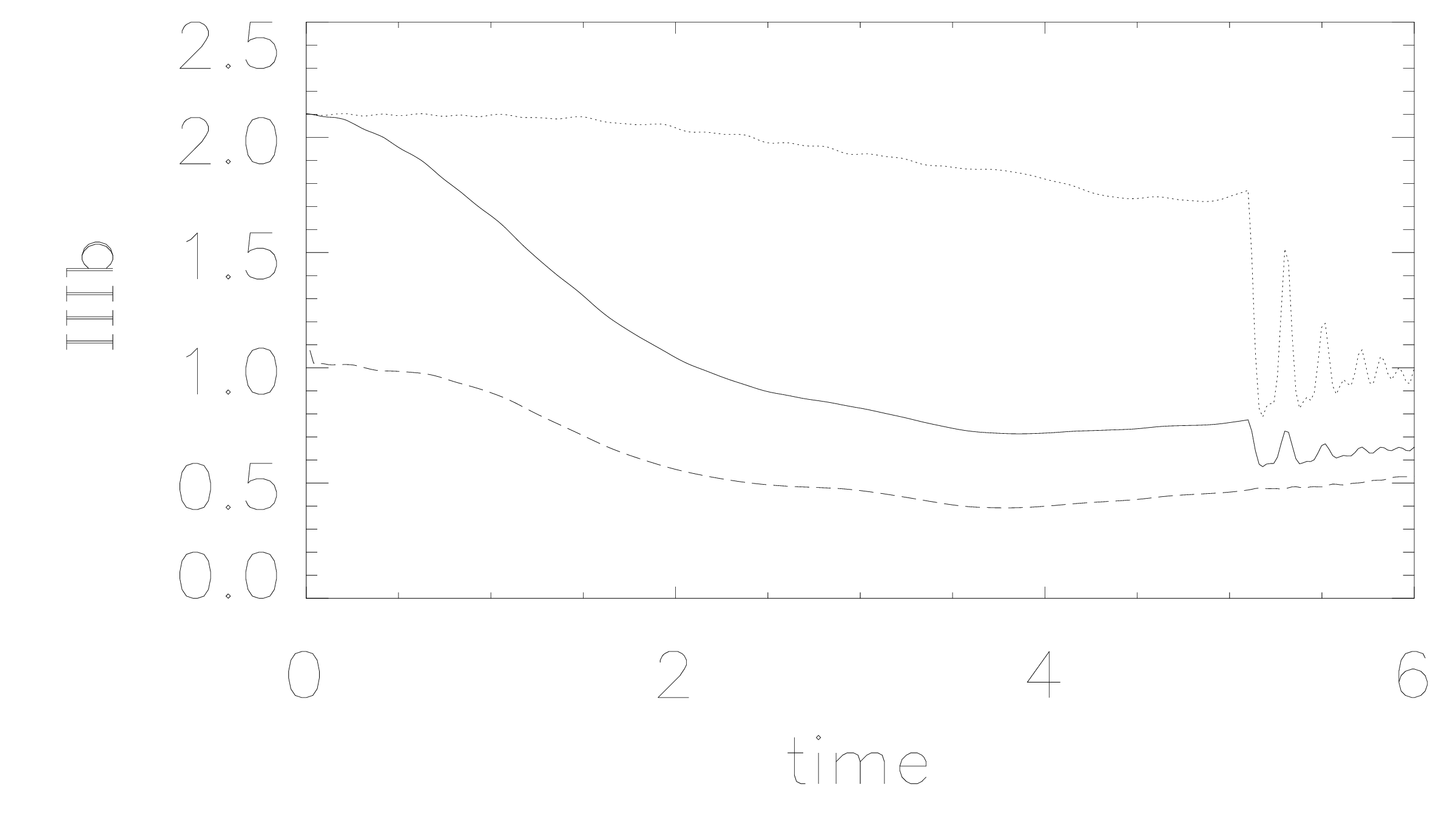}}
\caption{Temporal evolution of Alfv\'en ratios $r^A(t)$ (solid), $r^A_{2D}(t)$ (dash) and $r^A_w(t)$ (dot) for 
run ${\bf IIIb}$. 
\label{Fig++}}
\end{figure}
Figure~\ref{Fig++} shows eventually the time variations of the Alfv\'en ratios $r^A(t)$, $r^A_{2D}(t)$ and 
$r^A_w(t)$. The main difference between run ${\bf IIIa}$ (see Fig. \ref{FigtmprapAkb0}) and run ${\bf IIIb}$ 
resides first in the initial values which are larger for the latter run. The evolution is also different with a variation 
of all quantities during the driving phase. It is only during the decay phase that we recover the same behavior 
as in run ${\bf IIIa}$ with alfv\'enic fluctuations and an equipartition for the 3D modes. In this decay phase, 
a large deviation of $r^A_w(t)$ is observed immediately after $t^*$ in order to reach the equipartition which 
clearly means that it is the natural state of the system (in ${\bf IIIa}$ the situation was not totally clear since
before $t^*$ we were already at equipartition).

\section{Summary and conclusion}
\label{sec6}

In this paper, we present a set of 3D direct numerical simulations of incompressible driven MHD turbulence 
under the influence of a uniform magnetic field ${\bf B_0}$. In particular, the temporal and spectral properties 
of the 2D state (or slow mode) 
are investigated. We show that if initially the energy contained in the 2D state is set to zero it 
becomes shortly non negligible in particular when the intensity of $B_0$ is strong. For our larger $B_0$ 
intensity ($B_0=15$) this energy saturates in the stationary phase around $2/3$ of the total energy whereas 
the energy of the 3D modes remains roughly constant. We also observed that the ratio $B_0/b_{rms}$ 
saturates around $7$ for both simulations with $B_0=15$ with initially a ratio equal to $15$ (run ${\bf IIIa}$) 
and to $19$ (run ${\bf IIIb}$). In all situations, the magnetic energy dominates the kinetic energy but it is 
shown that at large $B_0/b_{rms}$ and in the decay phase the natural state for the 3D modes is the equipartition
whereas the 2D state is magnetically dominated. It is interesting to note that a theoretical model in terms of 
condensate has been recently proposed to explain the spontaneous generation of a residual energy 
$E^v - E^b$ at small $\kpa$ \citep{Boldyrev09}. This model is based on the breakdown of the mirror-symmetry 
in unbalanced wave turbulence. 

From a spectral point of view, when the $B_0$ intensity is strong enough the $\kpe$--energy spectra are mainly 
composed at large scales by the 2D state and at small scales by the 3D modes. This situation is similar to rotating 
turbulence for neutral fluids \cite{smith,baroud} where in particular the nonlinear transfers are reduced along the rotating rate \cite{Galtier03}. However, a detailed analysis of the temporal evolution of the 2D state energy spectra 
shows that a direct cascade happens whereas for rotating turbulence an inverse cascade is generally evoked
\citep{Mininni10}. The same 
remark holds for energy spectra at fixed $\kpa>0$ where we observe additionally a pinning effect at large $\kpe$. 
According to the value of $B_0/b_{rms}$ scalings close to $\kpe^{-3/2}$ -- $\kpe^{-5/3}$ or $\kpe^{-2}$ are found 
which are in agreement with different predictions for strong and wave turbulence \cite{GS95,ng96,Galtier2000,Boldyrev06}. 

The external force seems to be an important parameter for the dynamics. For example in \cite{Boldyrev08} a change 
of spectral slope was reported for the $\kpe$--energy spectrum when the intensity of $B_0$ is modified. In this work 
the external force was applied at $\kpa>0$ and $\kpe =1,2$. In \cite{Muller} a forcing which kept the ratio of 
fluctuations to mean field approximately constant was implemented by freezing modes $k \le 2$ and a spectral 
slope close $\kpe^{-3/2}$ was reported for $B_0=5$. The fact that the 2D state was imposed by the forcing has 
certainly an impact on the nonlinear dynamics since it prevents the natural growth of the 2D state energy at small 
$\kpe$. We remind that the 2D state is essential at large $B_0/b_{rms}$ since it mainly drives the nonlinear dynamics. 
In particular, we observe that a reduction of the 2D state energy of one type of polarity leads to a decrease of the 
inertial range of the 3D modes energy spectrum with the opposite polarity.

\begin{acknowledgments}
High performance computing resources were provided by Cubby, Nersc and Cicart. A. Bhattacharjee 
is gratefully acknowledged. We also thank Y. Ponty, H. Politano and Institut universitaire de France
for financial support. 
\end{acknowledgments}



\begin{thebibliography}{99}
\bibitem{elmegreen}
J. Scalo, and B.G. Elmegreen,  ARA\&A {\bf 42}, 275 (2004).
\bibitem{Govoni}
F. Govoni, et al., Astron. Astrophys. {\bf 460}, 425 (2006).
\bibitem{galtier06}
S. Galtier, J. Low Temp. Phys. {\bf 145}, 59 (2006).
\bibitem{Galtier06aa}
S. Galtier, J. Plasma Phys. {\bf 72}, 721 (2006).
\bibitem{Galtier08a}
S. Galtier, Phys. Rev. E {\bf 77}, 015302(R) (2008).
\bibitem{MontgoTurner}
D. Montgomery, and L. Turner, Phys. Fluids {\bf 24}, 825 (1981).
\bibitem{Grappin83}
R. Grappin, J. Leorat, and A. Pouquet, Astron. Astrophys.  {\bf 126}, 51 (1983).
\bibitem{GS95}
P. Goldreich, and S. Sridhar, Astrophys. J. {\bf 438}, 763 (1995).
\bibitem{ng96}
C.S. Ng, and A. Bhattacharjee, Astrophys. J. {\bf 465}, 845 (1996).
\bibitem{Amitava}
A. Bhattacharjee, C.S. Ng, and Spangler, Astrophys. J. {\bf 494}, 409 (1998).
\bibitem{politano98}
H. Politano, A. Pouquet, and V. Carbone, Europhys. Lett. {\bf 43}, 516 (1998). 
\bibitem{Galtier99}
S. Galtier, E. Zienicke, H. Politano, and A. Pouquet, J. Plasma Phys. {\bf 61}, 507 (1999).
\bibitem{Cho00}
J. Cho, and E.T. Vishniac, Astrophys. J. {\bf 539}, 273 (2000).
\bibitem{Galtier2000}
S. Galtier, S.V. Nazarenko, A.C. Newell, and A. Pouquet, J. Plasma Phys. {\bf 63}, 447 (2000).
\bibitem{Galtier2002}
S. Galtier, S.V. Nazarenko, A.C Newell, and A.  Pouquet, Astrophys. J. {\bf 564}, L49 (2002).
\bibitem{Galtier2005}
S. Galtier, A. Pouquet, and A. Mangeney, Phys. Plasmas {\bf 12}, 092310 (2005).
\bibitem{Mininni05}
P.D. Mininni, A. Alexakis, and A. Pouquet, Phys. Rev. E {\bf 72}, 046302 (2005). 
\bibitem{Muller}
W.-C. M\"uller, and R. Grappin, Phys. Rev. Lett. {\bf 95}, 114502 (2005).
\bibitem{Boldyrev06}
S. Boldyrev, Phys. Rev. Lett. {\bf 96}, 115002 (2006).
\bibitem{Galtier06}
S. Galtier, and B.D.G. Chandran, Phys. Plasmas {\bf 13}, 114505 (2006).
\bibitem{Alexakis07a}
A. Alexakis, Astrophys. J. {\bf 667}, L93 (2007). 
\bibitem{Alexakis07b}
A. Alexakis, B. Bigot, H. Politano, and S. Galtier, Phys. Rev. E {\bf 76}, 056313 (2007). 
\bibitem{Bigot07a}
B. Bigot, S. Galtier, and H. Politano, Phys. Rev. Lett. {\bf 100}, 074502 (2008). 
\bibitem{Bigot08b}
B. Bigot, S. Galtier, and H. Politano, Phys. Rev. E {\bf 78}, 066301  (2008). 
\bibitem{Boldyrev08}
J.-C. Perez, and S. Boldyrev, Astrophys. J. {\bf 672}, L61 (2008).
\bibitem{Chandran08}
B.D.G. Chandran, Astrophys. J. {\bf 685}, 646 (2008).
\bibitem{Boldyrev09}
S. Boldyrev, and J.-C. Perez, Phys. Rev. Lett. {\bf 103}, 225001 (2009).
\bibitem{Galtier09}
S. Galtier, Astrophys. J. {\bf 704}, 1371 (2009). 
\bibitem{Matt09}
W.H. Matthaeus, S. Oughton, and Y. Zhou, Phys. Rev. E {\bf 79}, 035401(R) (2009). 
\bibitem{Mininni09}
P.D. Mininni, and A. Pouquet, Phys. Rev. E {\bf 80}, 025401(R) (2009). 
\bibitem{Muller10}
W.-C. M\"uller, and R. Grappin, submitted (2010). 
\bibitem{Podesta10}
J.J. Podesta, and A. Bhattacharjee, submitted (2010). 
\bibitem{iro}
P.S. Iroshnikov, Soviet Astron. {\bf 7}, 566 (1964).
\bibitem{Kraichnan65}
R.H. Kraichnan, Phys. Fluids {\bf 8}, 1385 (1965).
\bibitem{ponty}
Y. Ponty, J.-P. Laval, B. Dubrulle, F. Daviaud, and J.-F. Pinton, Phys. Rev. Lett. {\bf 99}, 224501 (2007). 
\bibitem{MBG03}
W.-C. M\"uller, D. Biskamp, and R. Grappin, Phys. Rev. E {\bf 67}, 066302 (2003).
\bibitem{smith}
L.M. Smith, and F. Waleffe, Phys. Fluids {\bf 11}, 1608 (1999). 
\bibitem{baroud}
C.N. Baroud, B.B. Plapp, Z.-S. She, and H.L. Swinney, Phys. Rev. Lett. {\bf 88}, 114501 (2002). 
\bibitem{Galtier03}
S. Galtier, Phys. Rev. E {\bf 68}, 015301 (2003). 
\bibitem{Mininni10}
P.D. Mininni, P. Dmitruk, W.H. Matthaeus, and A. Pouquet, arXiv:1005.1574v1 (2010). 
\end{thebibliography}
\end{document}